 \documentclass[final,5p,times,twocolumn,authoryear]{elsarticle}

\usepackage{amssymb}
\usepackage{lipsum}
\usepackage{amsmath}
\usepackage{algpseudocode}
\usepackage{algorithm}
\usepackage{orcidlink}
\usepackage{tikz}
\usepackage{newtxtext, newtxmath}
\journal{Astronomy $\&$ Computing}

\begin{document}

\begin{frontmatter}



\title{Nessie: A Rust-Powered, Fast, Flexible, and Generalized Friends-of-Friends Galaxy-Group Finder in R and Python.}


\author[first]{Trystan S. Lambert\orcidlink{0000-0001-6263-0970}}
\author[first]{A.S.G. Robotham\orcidlink{0000-0003-0429-3579}}
\author[second]{M. Bravo\orcidlink{0000-0001-5742-7927}}
\author[first]{C. del P. Lagos\orcidlink{0000-0003-3021-8564}}
\author[first]{R. Tobar\orcidlink{0000-0002-1052-0611}}
\author[first]{S. Driver\orcidlink{0000-0001-9491-7327}}
\author[third]{A. Aufan Stoffels d'Hautefort\orcidlink{0009-0003-9019-1380}}

\affiliation[first]{organization={International Centre for Radio Astronomy Research (ICRAR)},%Department and Organization
            addressline={M468, University of Western Australia, 35 Stirling Hwy, Crawley}, 
            city={Perth},
            postcode={6009}, 
            state={WA},
            country={Australia}}

\affiliation[second]{organization={Department of Physics \& Astronomy},%Department and Organization
            addressline={McMaster University, 1280 Main Street W}, 
            city={Hamilton},
            postcode={L8S 4M1}, 
            state={ON},
            country={Canada}}

\affiliation[third]{organization={Astronomy Centre},%Department and Organization
            addressline={University of Sussex}, 
            city={Brighton},
            postcode={BN1 9QH}, 
            state={Falmer},
            country={UK}}

\begin{abstract}
We introduce \texttt{Nessie}, a galaxy group finder implemented in \texttt{Rust} and distributed as both a \texttt{Python} and \texttt{R} package. \texttt{Nessie} employs the friends-of-friends (FoF) algorithm and requires only on-sky position and redshift as input, making it immediately applicable to surveys that lack a well-defined luminosity function. We implement several algorithmic optimizations—including binary search and k-d tree pre-selection—that significantly improve performance by reducing unnecessary galaxy pair checks.

To validate the accuracy of \texttt{Nessie}, we tune its parameters using a suite of GALFORM mock lightcones and achieve a strong Figure of Merit. We further demonstrate its reliability by applying it to both the GAMA and SDSS surveys, where it produces group catalogues consistent with those in the literature. Additional functionality is included for comparison with simulations and mock catalogues.

Benchmarking on a standard MacBook Pro (M3 chip with 11 cores) shows that version 1 of \texttt{Nessie} can process $\sim$1 million galaxies in  $\sim10$ seconds, highlighting its speed and suitability for next-generation redshift surveys.

\end{abstract}




\end{frontmatter}

%\tableofcontents



\section{Introduction}

It is well established that galaxies are not randomly distributed throughout the Universe. Instead, they cluster in a variety of environments—groups, clusters, filaments, and voids—that together form the \textit{cosmic web}, a vast large-scale structure that traces the underlying distribution of dark matter \citep{peebles1980,jarrett2004,libeskind2018}. Because galaxies are biased tracers of this dark matter, their three-dimensional distribution in redshift space readily reveals the geometry of the cosmic web, which has been clear even in the earliest wide-area spectroscopic surveys \citep[e.g. the CfA stick man;][]{davis1982, huchra1983}.

While this structure is visually evident, identifying it by eye is inherently subjective and non-reproducible. To address this, numerous computational methods have been developed to systematically identify structure in galaxy surveys, in particular galaxy-groups, such as halo-based group finders \citep[e.g.][]{yang2005,yang2007,tinker2021} who use a halo occupation model and mock catalogues from simulations to iteratively find galaxy groups assuming a mass-to-light ratio and Voronoi tessellation techniques \citep[e.g.][]{ramella2001, elyiv2009} which rely on geometrically identifying clusters based on their density above the background. However, among the most enduring and popular is the \textit{Friends-of-Friends} (FoF) algorithm \citep{huchra1982}, which links galaxies together based on pairwise distances. Unlike the halo-based algorithms, the FoF method isn't tied to any specific simulation or mock catalogue and doesn't need to assume a mass-to-light ratio. At the same time the FoF technique is much more physically motivated than the geometric Voronoi methods. Its simplicity, flexibility, and effectiveness have made it a staple tool for building galaxy group catalogues across many redshift surveys \citep[e.g.,][]{eke2004, crook2007, robotham2011, tempel2014, duarte2014, tempel2016, lambert2020}. 

Despite its popularity, the FoF algorithm is often re-implemented from scratch for each new survey or catalogue. These implementations are frequently fine-tuned to match specific survey geometries or selection functions, making cross-survey comparisons difficult and limiting reproducibility. While many versions of the FoF algorithm have been proposed—some offering algorithmic improvements—there remain very few readily available, general-purpose software packages for applying FoF to redshift surveys. This scarcity is surprising given FoF’s widespread use and long-standing role in extragalactic astronomy.

Looking ahead, the next generation of redshift surveys will dramatically increase the volume and density of galaxy data; the \textit{4-Metre Multi-Object Spectroscopic Telescope} \citep[4MOST;][]{dejong2012} is set to begin observations shortly, with several of its extragalactic surveys—such as the \textit{Wide Area VISTA Extragalactic Survey} \citep[WAVES;][]{driver2019} and the \textit{4MOST Hemisphere Survey} \citep[4HS;][]{taylor2023}—expected to contribute tens of millions of galaxy redshifts. While FoF remains conceptually simple, its na\"ive implementation scales as $\mathcal{O}(n^2)$, making it computationally expensive for these massive datasets. In this new era, it is no longer practical to continue building bespoke group-finders from scratch for each survey.

In this paper, we present a generalized, high-performance FoF implementation designed for modern redshift surveys. Our approach builds on previous adaptations of the FoF algorithm, in particular the version used in the \textit{Galaxy and Mass Assembly} (GAMA) survey \citep{robotham2011}, and incorporates several new algorithmic improvements that significantly reduce runtime and memory usage. Crucially, we also provide an open-source implementation---\texttt{Nessie}---available as both an \texttt{R} and \texttt{Python} package. Our goal is to provide a group-finding tool that is: 
\begin{enumerate}
    \item generalized and can run on a wide range of redshift surveys,
    \item computationally efficient and scalable, and
    \item user-friendly and accessible to the community.
\end{enumerate}

The structure of this paper is as follows: in Section~2, we outline the classical FoF algorithm, describe its GAMA implementation, and present the optimizations introduced in \texttt{Nessie}. In Section~3, we validate \texttt{Nessie} using mock catalogues originally used in \cite{robotham2011}, and compare the resulting group catalogues to both GAMA and the Sloan Digital Sky Survey \citep[SDSS;][]{kollmeier2019}, including a comparison with the group catalogue of \cite{tempel2017}. We discuss these results in Section~4 and conclude in Section~5, where we also outline future development directions.

Throughout this work, we adopt a flat $\Lambda$CDM cosmology with $H_0 = 70$ km s$^{-1}$ and $\Omega_m = 0.3$.

\section{Methods}

\subsection{The Friends-of-Friends algorithm}
The FoF algorithm is a popular method of determining galaxy groups and has proven to be exceptionally powerful in identifying them in redshift surveys since the construction of the CfA group catalogue \citep{huchra1982}. However, whilst the core idea of the FoF algorithm has been used across numerous redshift surveys with varying geometries and depth \citep[e.g.][]{eke2004, crook2007, knebe2011, robotham2011, tempel2014, tempel2016, lambert2020} the exact implementation has often differed. Since \texttt{Nessie} is the formalized successor to the group finder used in \cite{robotham2011}, we explain the core FoF algorithm, how this was implemented for the GAMA survey in \cite{robotham2011}, and finally the improvements made in \texttt{Nessie}.

\subsubsection{Classical implementation}
 The FoF algorithm constructs groups based on a percolation algorithm which associates galaxies as belonging to the same group if they are found close together in redshift space. What ``close together'' means practically has been a matter of debate for every rendition of the algorithm. In its simplest sense, the algorithm checks if two galaxies are within some pre-defined tolerances in both redshift space and projected angular distance: i.e., $\Delta v_{ij} < V_{\rm L}$ and $D_{ij}\left(\theta\right) < D_{\rm L}$ -- where $\Delta v_{ij} = |v_{i} - v_{j}|$ is the difference in velocity between galaxies $i$ and $j$, $V_{\rm L}$ and $D_{\rm L}$ are the linking velocity and linking projected distance respectively, and $D_{ij}\left(\theta\right)$ is the projected angular distance between galaxy $i$ and galaxy $j$. If galaxies $i$ and $j$ satisfy both criteria, then they are considered ``friends''. Once all friends are found for one galaxy, the algorithm continues by looking for friends around all those galaxies. When no more friends are found, the total collection of friends are the group. This core algorithm remains essentially unchanged among all renditions of the group finder. The choice of $V_L$ and $D_L$ are where FoF algorithms mainly differ.

 The simplest implementation of the algorithm would be to set both $V_L$ and $D_L$ to constant values. However, this ignores a variety of selection effects such as variation of the sampling of the galaxy luminosity function \citep{huchra1982}. Whilst some implementations of the algorithm, especially at low redshifts, can choose a constant value for $V_L$ \citep[e.g.][]{crook2007, lambert2020}, much more common is to scale both linking lengths. Classically, this is done using the sampled luminosity function and the overdensity factor above critical density.

\begin{equation}
    D_L = D_0 \left[\frac{\int\limits_{-\infty}^{M_{\rm lim}}\Phi\left(M\right)dM}{\int\limits_{-\infty}^{M_{ij}} \Phi\left(M\right)dM}\right]^{1/3},
\end{equation}
 \begin{equation}
    V_L = V_0 \left[\frac{\int\limits_{-\infty}^{M_{\rm lim}}\Phi\left(M\right)dM}{\int\limits_{-\infty}^{M_{ij}} \Phi\left(M\right)dM}\right]^{1/3},
\end{equation}

where $\Phi(M)$ is the differential luminosity function, $D_0$ is the projected separation at a fiducial redshift, and $V_0$ is some fiducial redshift difference. $M_{\rm lim}$ is the absolute magnitude of the apparent magnitude limit of the survey at this fiducial redshift and $M_{ij}$ is the absolute magnitude of the same apparent magnitude limit, but at the average redshift between galaxies $i$ and $j$. The choice of $D_0$ determines the density contrast of the identified groups \citep{crook2007}, and the ratio between $D_0$ and $V_0$ is related to the assumed cosmological mean density \citep{huchra2012}. 

\begin{equation}
\label{eq: density_constrast}
    \frac{\rho}{\bar{\rho}} = \frac{3}{4\pi D_0^3} \left[\int\limits_{- \infty}^{M_{\rm lim}}\Phi\left(M\right)dM\right]^{-1} -1,
\end{equation}

Although this is one method of scaling $D_L$ and $V_L$, this is not the only method. Indeed, choosing how $V_L$ and $D_L$ scale is the first choice to be made when implementing the FoF algorithm on redshift surveys, followed by how to choose the constants $V_0$ and $D_0$. For more discussion on the classical implementation of the FoF algorithm we point the reader to \cite{huchra1982} and \cite{crook2007}.

\subsubsection{The GAMA group finder}
\texttt{Nessie} is based on the particular implementation of the FoF algorithm used to identify galaxy groups in the GAMA survey \citep{robotham2011}. For the particular case of GAMA, the linking lengths were defined in comoving space as opposed to velocity space as in \cite{huchra1982} but follows the core algorithm. Namely:

\begin{equation}
    \tan\left(\theta_{ij}\right)\left[\frac{D_{c,i} + D_{c,j}}{2}\right] \leq D_L,
\end{equation}

and 
\begin{equation}
|D_{c, i} - D_{c,j}| \leq V_L,
\end{equation}

where $D_c$ represents the comoving distance and $\theta_{ij}$ is the angular separation between galaxies $i$ and $j$. Unlike \cite{huchra1982}, $V_L$ and $D_L$ are coupled together via $V_L = R_{ij}D_L$, i.e., the average line-of-sight linking length between galaxies $i$ and $j$ ($V_{L}$) is related to the average plane-of-sky linking length between galaxies $i$ and $j$ ($D_L$) through some factor $R_{ij}$.

Originally, in \cite{robotham2011}, the linking lengths for galaxy $i$, were defined to be:

\begin{align}
    D_i &= \frac{b_iL_i}{c_i^{1/3}} \\
    V_i &= \frac{b_iR_iL_i}{c_i^{1/3}},
\end{align}

where $c_i$ is the completeness around galaxy $i$ as defined by the user, and $L_i$ is defined as:
\begin{equation}
\label{eq: core linking length}
    L_i = \left[\int\limits_{-\infty}^{M_{\rm lim, i}}\Phi\left(M\right)dM\right]^{-1/3} \left(\frac{\Phi\left(M_{\rm lim, i}\right)}{\Phi\left(M_{i}\right)}\right)^{\nu/3},
\end{equation}

and $b_i,$ and $R_i$ are defined as:

\begin{align}
    b_i & = b_0\left(\frac{1}{\Delta} \frac{\rho_{i}}{\bar{\rho}(z_i)}\right)^{E_b}, \\
    R_i & = R_0\left(\frac{1}{\Delta} \frac{\rho_{i}}{\bar{\rho}(z_i)}\right)^{E_R}.
\end{align}
The integral in the first part of Eq. \ref{eq: core linking length} is related to Eq. \ref{eq: density_constrast}, but the choice was made to scale this by the ratio of the luminosity function at the effective absolute magnitude limit of the survey at the position of galaxy $i$, $\Phi(M_{\rm lim, i})$, and at the absolute magnitude of the same galaxy, $\Phi(M_i)$. This allowed for larger linking lengths for galaxies which were intrinsically brighter. The strength of this scaling is determined by the free parameter $\nu$.

Both $b_i$ and $R_i$ rely on several free parameters which need to be fit, including $b_0$, $R_0$, $E_b$, $E_R$, and the density contrast $\Delta$. Furthermore, both rely on $\rho_i$, the empirical density which is calculated from the number of galaxies within a comoving cylinder with a radius of $\Delta_r$ and a length of $\Delta_l$ where $\Delta_l$ is scaled to account for finger-of-god effects (the apparent elongation of the 3D distribution of galaxies along the line of sight in redshift space due to peculiar velocities): $\Delta_l = K\Delta_r$. This also requires constructing $\bar{\rho}(z_i)$ -- the average local density implied by the $n(z)$ distribution, at the redshift of galaxy $i$.

All together, this original definition, used for versions 1 through 5 of the GAMA group catalogue, relies on eight free parameters which can be tuned: $b_0, R_0, \nu, E_b, E_R, \Delta, \Delta_r$ and $K$. Although it should be noted that $\Delta, \Delta_r$ and $K$ were all fixed in \cite{robotham2011}, so only five parameters were tuned.

Versions 6 through 10 of the GAMA group catalogue, redefine $L_i$ as

\begin{equation}
    L_i \equiv \bar{\rho}(z_i)^{-1/3} \left(\frac{\Phi\left(M_{\rm lim, i}\right)}{\Phi\left(M_{i}\right)}\right)^{\nu/3}.
\end{equation}

Replacing the integral term in Eq. \ref{eq: core linking length} with $\bar{\rho}(z)$ allows for more accuracy across a wider range of redshifts and can be built directly from the redshift distribution of the data itself. There are several ways to calculate $\bar{\rho}(z)$ since there are numerous ways to estimate the $n(z)$ distribution, but in the case of the GAMA survey this function was built by applying a KDE estimate from a random catalogue built specifically for GAMA-III by \cite{farrow2015}. These random catalogues are built to mimic the $n(z)$ distribution of an infinitely large magnitude limited survey without clustering \citep{cole2011}, effectively removing large-scale structure inhomogeneities. 

A probability density function (PDF) is then estimated from this random, unclustered, distribution of galaxies assuming a rectangular kernel with bandwidth $\sigma = \Delta x/\sqrt{12}$. Where $\Delta x$ is bin-width. This can then be used to determine the integral of bin $k$:
\begin{equation}
\label{eq: integral}
    I_k = \int\limits_{x_k - \Delta x/2}^{x_k+\Delta x/2}PDF(x)dx,
\end{equation}

and the mean density for an individual bin $i$ can be calculated as:
\begin{equation}
\label{eq: bins}
    \bar{\rho}_k = \frac{N_k I_k}{V_k},
\end{equation}
where $N_k$ is the number of random galaxies in bin $k$ and $V_k$ is the volume of that same bin. We finally interpolate over all the bins to obtain the functional form of $\bar{\rho}(z)$.

\subsubsection{The FoF implementation in \texttt{Nessie}}
Whilst the current implementation of the GAMA group catalogue has proven incredibly reliable, it requires fitting five free parameters, which need to be chosen in some way. And unlike the classical implementation which only need reasonable choices for two constants, that can be determined in a variety of ways (including mapping out the entire parameter space), this method can only be used by tuning these numerous free parameters using a simulation. Comparing to these mock catalogues is firstly computationally expensive and requires several days using high performance computing, and secondly, can tie the choice of parameters to a single simulation. Nevertheless, this is still a completely valid (and arguably best) method for determining these constants. 

However, when this tuning was done, \cite{robotham2011} noted that the choice of $b_0$ and $R_0$ dominated over all the other free parameters. In particular $E_b$ and $E_R$ had optimal values of $0.00$ and $-0.02$ (consistent with 0 within errors) respectively. Therefore, they themselves suggest that these parameters might be removed in future versions of the group catalogue. Therefore, we choose to set $E_b = E_R =0$, which then implies that $b_i = b_0$ and $R_i = R_0$. Moreover, $\Delta, \Delta_r,$ and $K$ have no effect when $E_b = E_R = 0$. This small and justifiable decision results in having to choose only three free parameters. 

In addition, even though $\nu$ was determined to be $0.63$ in \cite{robotham2011}, its practical effect on $L_i$, we argue, is negligible. Indeed, this was found to be the least impactful free-parameter in \cite{robotham2011}. Across the GAMA sample, the median value for $\left(\Phi(M_{\rm lim, i})/\Phi(M_i)\right)^{\nu/3}$ is $1.1 \pm 0.4$. And while this does affect some galaxies' $L_i$ value, the final linking length between two galaxies is the mean between them, mitigating this effect. Furthermore, galaxies which do receive a significant boost in their $L_i$ are already large, bright galaxies which tend to be in over-dense regions anyway, which means changing in $L_i$ has very little practical effect in determining galaxy groups. And finally, $\nu$ is not independent of $R_0$ and $b_0$ and any overall scaling would be absorbed in these two parameters. Therefore, we also choose to set $\nu =0$.

By setting $E_b = E_R = \nu = 0$ we reduce the number of possible free parameters from eight to two. Not only does this greatly cut down on computational time for comparisons to simulations, it also simplifies the overall algorithm tremendously. The plane-of-sky and line-of-sight linking lengths for galaxy $i$ are now respectively defined as:
\begin{align}
    D_i &\equiv b_0 \left[\frac{1}{\bar{\rho}(z_i)c_i}\right]^{1/3}, \\
    V_i &\equiv R_0D_i,
\end{align}

  The completeness $c_i$ can be calculated in a number of ways, however, we do provide a helper method, \texttt{calculate$\_$completeness} which takes the target RA and Dec coordinates of the survey as well as some user-defined radii, and counts the number of galaxies that were observed within the radius of a given galaxy, and the number of galaxies targeted within the same radius. The ratio of these two terms gives us a completeness between 0-1 for every galaxy in the redshift catalogue. This is done very quickly by using a k-d tree nearest neighbour search algorithm using the \texttt{kiddo} crate in \texttt{Rust}.

After making the above simplifications the full linking length equations for \texttt{Nessie} simplify to:

\begin{equation}
\label{eq: pos}
    \tan\left(\theta_{ij}\right)\left(D_{c,i} + D_{c,j}\right) \leq D_i + D_j,
\end{equation}

and 
\begin{equation}
\label{eq: los}
|D_{c, i} - D_{c,j}| \leq \frac{V_i + V_j}{2}.
\end{equation}

It is worth noting that Equations \ref{eq: pos} and \ref{eq: los} are entirely independent of $\Phi(M)$ and only rely on the three core properties that are available in a redshift survey, namely RA, Dec, and redshift, making it much more generalized over the classic version which at a minimum required a choice of luminosity function. 

We summarize the list of free parameters that have been chosen in different group finders in Table \ref{tbl: parameter}.

\begin{table*}
    \caption{Table showing the various free parameters that have been adopted through multiple different group finders, along with brief explanations and references to examples where they were used.}
    \label{tbl: parameter}
\begin{tabular}{lll}
\hline \hline
Free Parameter & Explanation                                                                     & Used in                                                                                                    \\ \hline
$V_0$          & Fiducial velocity difference                                                    & \cite{huchra1982}, \cite{crook2007}, \cite{lambert2020} \\
$D_0$          & Projected separation constant                                                   & \cite{huchra1982}, \cite{crook2007}, \cite{lambert2020} \\
$b_0$          & Same as $D_0$ but in comoving space                                             & \cite{robotham2011}, This work                                                            \\
$R_0$          & Factor by which to scale $b0$ & \cite{robotham2011}, This work                                                            \\
$E_b$          & Plane-of-sky exponential scaling factor                 & \cite{robotham2011}, \cite{eke2004}                                      \\
$E_R$          & Line-of-sight exponential scaling factor                & \cite{robotham2011}, \cite{eke2004}                                      \\
$\nu$          & Softening power                                                                 & \cite{robotham2011}                                                                       \\
$\Delta$  & Density Contrast                                                               & \cite{robotham2011}                                                                       \\
$\Delta_r$     & Radius of comoving cylinder                      & \cite{robotham2011},                                                                      \\ 
$K$            & Factor by which to scale $\Delta_r$ ($\Delta_l$ = K $\Delta_r$)                 & \cite{robotham2011}    \\             \hline                                                     
\end{tabular}
\end{table*}
An important element which existed in the GAMA group catalogue, but was never discussed in \cite{robotham2011} is the idea of maximum allowable linking lengths. In order to avoid runaway, where the linking lengths are non-physically large (e.g. if $\bar{\rho}(z)$ is very near zero). We choose to set these maximum line-of-sight and plane-of-sky linking lengths, to be the redshift dispersion and comoving angular virial radius for a galaxy cluster with a mass of $10^{15}M_{\odot}$, which is the largest expected cluster mass. Links larger than these are certainly non-physical. We use the equations for the velocity dispersion, and virial radius as a function of $\Delta_{\rm vir}$ from \cite{Robotham2016}, namely:
\begin{equation}
\label{eq: max pos}
    D_{\rm max}(z) = \left(\frac{3M_{\rm max}}{4\pi \Delta_{\rm vir}\rho_{c}(z)}\right)^{1/3},
\end{equation}

\begin{equation}
\label{eq: max los}
    V_{\rm max}(z) = M_{\rm max}\frac{(1+z)}{H(z)}\sqrt{\frac{32}{3} \pi G^3\Delta_{\rm vir}\frac{\rho_{c}(z)}{(1+z)^3}},
\end{equation}

where $\rho_{c}(z)$ is the critical energy density of the Universe at a redshift of $z$ -- the energy density needed for a flat universe \citep{hamilton2001}
\begin{equation}
    \rho_{c}(z) = \frac{3H(z)^2}{8\pi G} \left[\frac{M_{\odot}}{\text{Mpc}^3}\right],
\end{equation}
$G$ is the gravitational constant in units of ($M_{\odot} \text{Mpc}^{-1} \text{km}$ $\text{s}^{-1}$), $H(z)$ is the Hubble constant at a given redshift \citep{peebles1980}
\begin{equation}
    H(z) = H_0\sqrt{\Omega_m(1+z)^3 + \Omega_k (1+z)^2 + \Omega_{\Lambda}},
\end{equation}
where $\Omega_m$, $\Omega_k$, and $\Omega_{\Lambda}$ are the matter density parameter, curvature density parameter, and dark energy density parameter respectively \citep{peebles1980}. The max viral mass is taken to be $M_{\rm max} = 10^{15}M_{\odot}$, and $\Delta_{\rm vir}$ is chosen to be 200. 

Any linking length which exceeds these maximum values are assigned, instead, the maximum values themselves.

The maximum allowable linking lengths (Equations \ref{eq: max pos} and \ref{eq: max los}) are much larger than the mean linking length between galaxies. It ensures that any linking lengths which are greater than these limits are non-physical, since it constrains inter-galaxy links to within the radius of a large cluster. In addition, imposing these limits also allows us to make certain speed-ups discussed later on.

\subsection{Comparing to mock catalogues}
As mentioned, one method of choosing the constants $b_0$ and $R_0$ is to run the group finder on a mock catalogue with known groupings and then search for the combinations of free parameters that result in the largest score metric. \texttt{Nessie} has this ``tuning'' functionality built in, although this is not necessary to the core group finding algorithm. 

What metric to optimize can itself be a contentious issue. However, we choose to follow the method of tuning described in \cite{robotham2011} exactly. Therefore, we point the reader to section 3.1 of that paper for a full discussion but include the equations here (in a more compact form) for completeness. 

The core score metric is dependent on grouping efficiency ($E_{\rm tot}$) and grouping purity ($Q_{\rm tot}$). In the case of grouping efficiency, this is simply the product of the number of bijective groups divided by the total number of groups in each catalogue:
\begin{equation}
E_{\rm total} = E_{\rm FoF}E_{\rm mock} = \left(\frac{N_{\rm bij}}{N_{\rm FoF}} \right) \left(\frac{N_{\rm bij}}{N_{\rm mock}}\right),
\end{equation}
where $N_{\rm FoF}$, $N_{\rm mock}$, $N_{\rm bij}$  are the number of groups that are found using the group-finder, the number of ``true'' groups according to the mock catalogue, and  are the number of bijective groups (defined to be groups that share more than $50\%$ of their members between catalogues) respectively.

The total grouping purity is the product of the grouping purity in FoF and mock catalogues --  $Q_{\rm total} = Q_{\rm FoF}Q_{\rm mock}$. For each catalogue, the group purity is defined as
\begin{equation}
    Q_{\rm FoF} = \frac{\sum\limits_{i=1}^{N_{\rm FoF}}(P_{\text{FoF},i})(n_{\text{FoF},i})}{G_{\rm FoF}},
\end{equation}

and 

\begin{equation}
    Q_{\rm mock} = \frac{\sum\limits_{i=1}^{N_{\rm mock}}(P_{\rm mock,i})(n_{\text{mock},i})}{G_{\rm mock}},
\end{equation}
where $G_{\rm FoF}$ and $G_{\rm mock}$ are the total number of galaxies in groups in the FoF and mock catalogues respectively, $n_{\text{FoF},i}$ and $n_{\text{mock},i}$ are the number of galaxies in group $i$ in the FoF and mock catalogues respectively, and $P_{\rm FoF}$ and $P_{\rm mock}$ are the purity products of the $i^{\rm th}$ best-matching FoF and mock group.

The total score we choose to maximize is defined as 
\begin{equation}
    S_{\rm total} = E_{\rm total} Q_{\rm total},
\end{equation}

and ranges from 0 to 1. However, it is important to note that the $E_{\rm total}$ and $Q_{\rm total}$ are both the product of two ratios, with the FoF and mock catalogues both contributing. This implies that $S_{\rm total}$ is made up of the product of four ratios, and so the average contribution of those ratios scales as $S_{\rm total}^{1/4}$. For example, a score of $S_{\rm total} = 0.3$ implies that the average value of $E_{\rm mock}$, $E_{\rm FoF}$, $Q_{\rm mock}$, $Q_{\rm FoF}$ is $0.74$. 

Often, when comparing to a mock catalogue, it is useful to compare to multiple instances to account for cosmic variance. \texttt{Nessie} calculates the average score among multiple different mock catalogues by using the harmonic mean which is most appropriate for averaging over ratios. 

\subsection{Calculating the galaxy to galaxy links}
Choosing the correct scaling of linking lengths, and appropriate methods for determining the free parameters of those linking lengths are the most complex part of implementing a FoF algorithm. However, this doesn't actually represent the most computationally intensive aspect. Once the linking lengths have been defined, it is relatively inexpensive to generate. Much more intensive is determining the links between individual galaxies which is largely driven by physically determining the on-sky and line-of-sight separations. 

In the worst case, each galaxy $i$ has to determine if it is linked to every galaxy $j$ which scales as $\sim \mathcal{O}(n^2)$. This scaling makes writing the core FoF algorithm incredibly slow in dynamic languages such as \texttt{R} or \texttt{Python}. For example, \texttt{pyFoF}, a pure \texttt{Python} implementation of the classic FoF algorithm \footnote{\url{https://github.com/TrystanScottLambert/pyFoF/}} processes $\sim 50 000$ galaxies in about 10 minutes. This is not sufficient with the next generation of redshifts surveys to begin observations within the next year and promising to deliver millions of redshifts \citep[e.g.][]{driver2019, taylor2023}. 

We instead build the main link-finding functionality in \texttt{Rust}, an extremely fast and memory safe compiled language. This core functionality is published to the \texttt{Rust} package environment (Cargo), as \texttt{fof}\footnote{\url{https://crates.io/crates/FoF}}. We then build the \texttt{R} and \texttt{Python} packages by wrapping the core functionality of the \texttt{fof} crate using the \texttt{rextendr}\footnote{\url{https://cran.r-project.org/web/packages/rextendr/index.html}} and \texttt{pyO3}\footnote{\url{https://docs.rs/pyo3/latest/pyo3/}} packages for \texttt{R} and \texttt{Python} respectively. Firstly, this allows for other projects to lean on the core link-finding functionality in their own group-finder implementations. And secondly, this ensures that updates to the \texttt{fof} \texttt{Rust} code will automatically be included in both \texttt{python} and \texttt{R} packages at the same time.

This design choice already makes the algorithm faster as compared to a pure \texttt{Python} or \texttt{R} solution simply by using a compiled language over interpreted ones. Indeed, \cite{robotham2011} wrote the core FoF algorithm in \texttt{Fortran} and \texttt{c++} and other group finders have relied on compiled languages for speed boosts. However, we further improve \texttt{Nessie}'s speed by simplifying the computational complexity of the most called functions, reducing the overall number of calculations, and multithreading the entire program.

\subsubsection{Fast searching along line-of-sight direction}
Even if we optimize the base functionality to be as fast as possible, it stands to reason that running any function less will result in less time. For the FoF algorithm this can be done by reducing the actual number of times galaxies need to be tested to be friends.

If we assume that the FoF algorithm is commutative, that is to say that if galaxy $i$ is a friend of galaxy $j$, then there is no reason to double-check galaxies. This implies that galaxy $i$ need only check galaxy $j$ where $j>i$. This algorithm is written explicitly in Alg. \ref{algo: 1}.

\begin{algorithm}
\caption{Commutative FoF checks}
\label{algo: 1}
\begin{algorithmic}[1]
\State Initialize \textit{link\_list} as empty
\For{$i = 0$ \textbf{to} $n-1$}
    \For{$j = i+1$ \textbf{to} $n$}
        \If{\Call{is\_friend}{$i, j$}}
            \State Append $(i, j)$ to \textit{link\_list}
        \EndIf
    \EndFor
\EndFor
\end{algorithmic}
\end{algorithm}

This drops the number of times we check for friends from $n^2$ to $n(n-1)/2$, essentially populating a strict upper triangular matrix versus the full square $n\times n$ matrix. However, while this is better (and all FoF implementations should at least loop over the candidates like this) this algorithm still has $\mathcal{O}(n^2)$ complexity.

We improve on this by restricting the number of friend-checks to only galaxies that meet the maximum line-of-sight linking length defined in Eq. \ref{eq: max los}. We can define the largest maximum line-of-sight linking length $V^*$ as:
\begin{equation}
    V* \equiv \max\limits_{i \in\{0, \dots ,n\}} \left\{V_{\rm max}(z_i) \mid \forall i, V_{\rm max}(z_i) \leq V^* \right\}.
\end{equation}

Where $n$ is the number of galaxies in the redshift survey. We can then guarantee that for any galaxies $i$ and $j$, if $|D_{c,i} - D_{c,j}| > (V^* + V_i)/2$ then galaxy $i$ cannot be friends with galaxy $j$, irrelevant of how close their angular separation is. This means the algorithm can be made faster by only performing the expensive friend-check for galaxies $i$ and $j$ where
\begin{equation}
\label{eq: search}
    |D_{c,i} - D_{c,j}| \leq \frac{(V^* + V_i)}{2}, \quad \forall i \in \{0, \dots, n-1\},\ \forall j > i \}.
\end{equation}

Most modern group finders will apply the line-of-sight condition first (e.g. \cite{robotham2011, lambert2020}) but \texttt{Nessie} does this very quickly by first sorting the comoving distances and then applying a binary search algorithm. This is even faster considering we need only search values greater than $i$. This allows \texttt{Nessie} to scale much better than $\mathcal{O}(n^2)$ as $n$ increases along the radial direction, i.e., as the survey increases with redshift. 

\subsubsection{Fast searching across plane-of-sky direction}

Similarly to the line-of-sight case, we aim to construct a fast spatial search in the plane-of-sky direction to restrict the computationally expensive “friend” check to only those galaxies within a maximum allowed angular separation. Unlike the line-of-sight direction, however, this task is more complex, since the plane-of-sky search is inherently two-dimensional, requiring the evaluation of angular separations over both right ascension (RA) and declination (Dec). This typically involves computationally expensive trigonometric operations.

To accelerate this process, we can transform the angular search into an equivalent problem in three-dimensional Cartesian space by working with chord distances. Instead of identifying all galaxies \( j \) such that the angular separation from galaxy \( i \), \( \theta_{ij} \), satisfies \( \theta_{ij} < \theta_{\rm max} \), we instead search for all galaxies whose chord distance from \( i \) is less than a corresponding threshold:
\begin{equation}
    \| \mathbf{x}_i - \mathbf{x}_j \| \leq d_{\rm max},
\end{equation}

where \( \mathbf{x}_i = (x_i, y_i, z_i) \) and \( \mathbf{x}_j = (x_j, y_j, z_j) \) are the unit vectors pointing to galaxies \( i \) and \( j \), obtained by converting RA and Dec into Cartesian coordinates on the unit sphere. The chord distance between two points on the unit sphere is related to their angular separation \( \theta_{ij} \) via:

\begin{equation}
    \| \mathbf{x}_i - \mathbf{x}_j \| = 2\sin\left( \frac{\theta_{ij}}{2} \right).
\end{equation}

Therefore, the condition \( \theta_{ij} < \theta_{\rm max} \) is equivalent to:

\begin{equation}
    \| \mathbf{x}_i - \mathbf{x}_j \| < 2\sin\left( \frac{\theta_{\rm max}}{2} \right) \equiv d_{\rm max}.
\end{equation}

Because the function \( 2\sin(\theta/2) \) is strictly increasing for \( \theta \in [0^\circ, 180^\circ] \), this transformation preserves the ordering of separations. That is, the set of galaxy pairs satisfying the angular constraint is exactly the same as the set satisfying the corresponding chord-distance constraint. Since we are only interested in whether the separation falls below the threshold (and not in the exact value), this allows us to replace the angular condition with a purely Euclidean one without loosing any accuracy.

In our algorithm, \( d_{\rm max} \) corresponds to the maximum projected linking distance \( D_{\rm max}(z) \), as previously defined in Eq.~\ref{eq: max pos}. Moreover, the left-hand side of Eq.~\ref{eq: pos} effectively represents the projected 3D separation between galaxies \( i \) and \( j \), which is the straight-line Euclidean distance in Cartesian space.

This transformation allows us to construct a k-d tree in 3D Euclidean space to efficiently identify all galaxies within a specified distance of a given galaxy. Querying a k-d tree has an average time complexity of \( \mathcal{O}(n \log n) \). However, there is an associated overhead in constructing the tree, and in this specific context, querying the k-d tree is still slower than a simple binary search along the line-of-sight. Therefore, minimizing the number of 3D spatial searches is crucial for optimal performance.

To accomplish this, we first restrict candidate galaxies \( j \) using the line-of-sight criterion defined in Eq.~\ref{eq: search}, and then use this subset to determine a tailored search radius for each galaxy \( i \):
\begin{equation}
    d_{{\rm max}, i} = \max \left\{ D_j \;\middle|\; \left| D_{c,i} - D_{c,j} \right| \leq \frac{V^* + V_i}{2} \right\}.
    \label{eq: max_pos_possible}
\end{equation}
In the worst case, \( d_{{\rm max}, i} \) equals the global maximum \( D_{\rm max}(z) \), but in practice it is often significantly smaller. For a galaxy $j$ to be considered it must therefore satisfy
\begin{equation}
\label{eq: pos search}
    \| \mathbf{x}_i - \mathbf{x}_j \| \leq d_{{\rm max}, i}.
\end{equation}

This means that galaxy \( j \) must satisfy both Eqs. \ref{eq: max los} and \ref{eq: max_pos_possible} to be considered a potential “friend” of galaxy \( i \) which ensures three conditions:
\begin{enumerate}
    \item \( j > i \),
    \item \( j \) is within the maximum possible line-of-sight distance, and
    \item \(j\) is within the maximum possible plane-of-sky distance.
\end{enumerate}

This subset of galaxies can still have galaxies that are not friends of $i$ but all of $i$'s friends are contained within the set that meet those criteria. Furthermore, finding the galaxies that satisfy Eqs.~\ref{eq: search} and~\ref{eq: pos search} can be performed efficiently using a combination of a binary search (for the line-of-sight criterion) and a k-d tree query (for the projected separation). This filtering leaves only a few candidate galaxies that require explicit friend checking.

This is a marked improvement upon the symmetric algorithm described in Algorithm~\ref{algo: 1}, significantly reducing the computational burden. The final linking procedure implemented in \texttt{Nessie} is shown in Algorithm~\ref{algo: improved algo}.

\begin{algorithm}
\caption{Optimized FoF checks with LoS and projected distance constraints}
\label{algo: improved algo}
\begin{algorithmic}[1]
\State Initialize \textit{link\_list} as empty
\State Sort galaxies by comoving distance \( D_c \)
\State Build k-d tree from 3D Cartesian coordinates \( \mathbf{x}_j = (x_j, y_j, z_j) \)
\For{$i = 0$ \textbf{to} $n-1$}
    \State Find all $j$ such that $j > i$ and \( \left| D_{c,i} - D_{c,j} \right| \leq \frac{V^* + V_i}{2} \) using binary search
    \State Set \( d_{{\rm max}, i} = \max \left\{ D_j \;\middle|\; \left| D_{c,i} - D_{c,j} \right| \leq \frac{V^* + V_i}{2} \right\} \)
    \State Use k-d tree to find all $j$ within \( d_{{\rm max}, i} \) of \( \mathbf{x}_i \)
    \For{each $j$ satisfying both criteria}
        \If{\Call{is\_friend}{$i, j$}}
            \State Append $(i, j)$ to \textit{link\_list}
        \EndIf
    \EndFor
\EndFor
\end{algorithmic}
\end{algorithm}

Algorithm \ref{algo: improved algo} has sub-quadratic but super-linear complexity. However, it does guarantee the exact same behaviour as Algorithm \ref{algo: 1} with much lower time complexity. Indeed, in the \texttt{fof} package we run an integration test which reads in the GAMA surveys and runs both algorithms and only passes if every link is exactly the same.

\begin{figure*}
    \centering
    \includegraphics[width=0.47\linewidth]{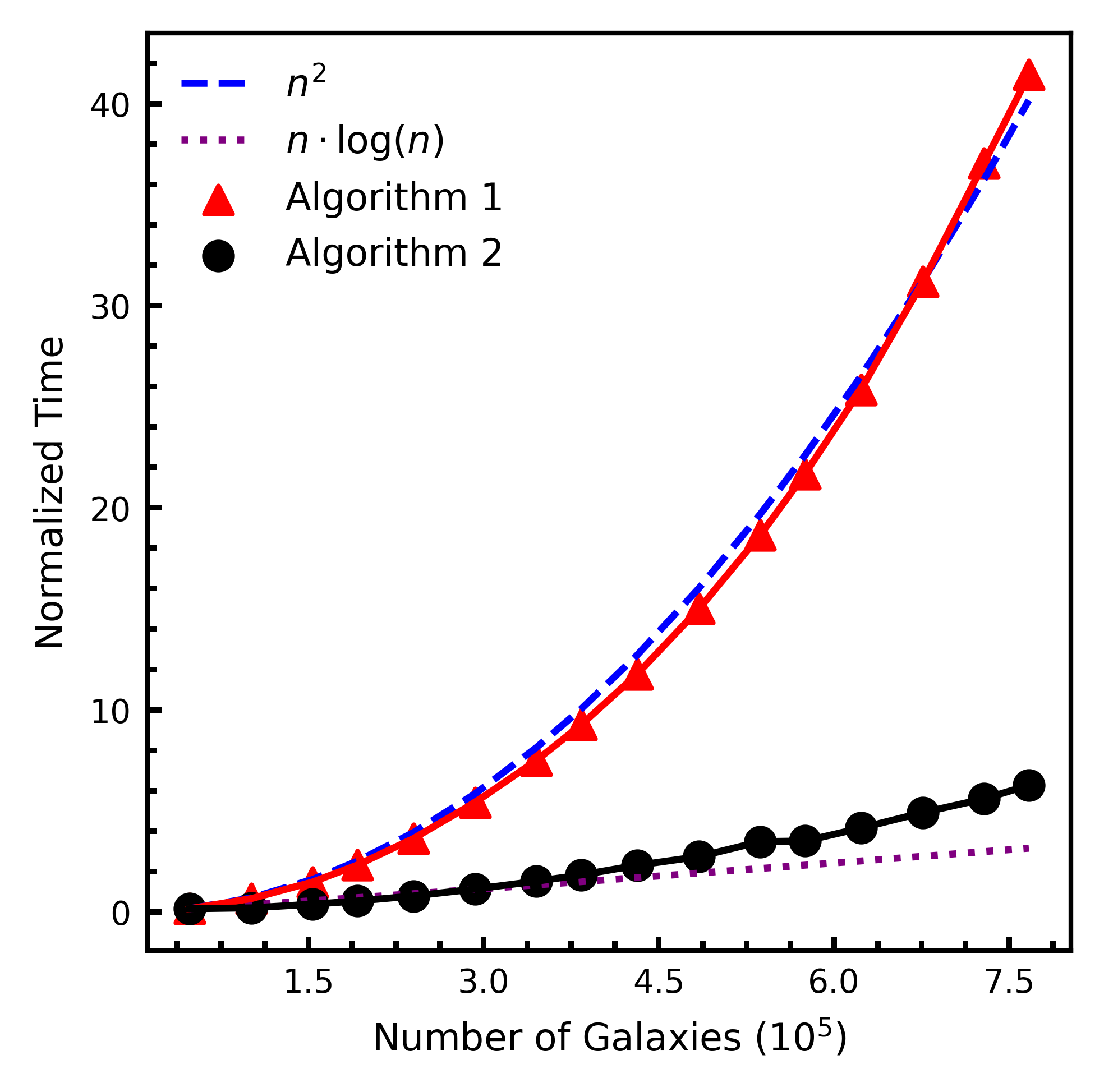}
    \includegraphics[width=0.47\linewidth]{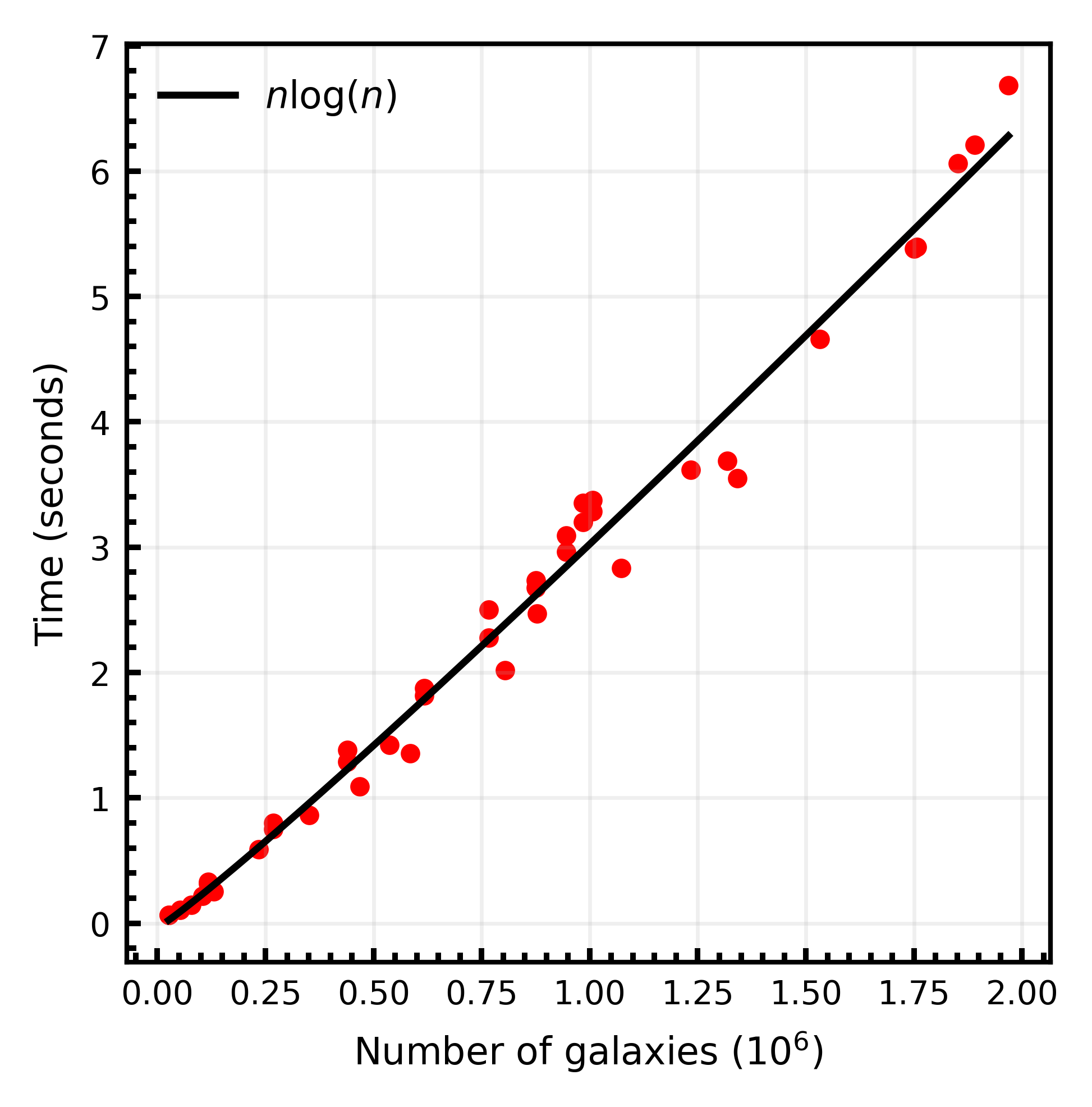}
\caption{
    Scalability of the group-finding algorithms. \textbf{Left panel:} Comparison of Algorithm~1 and Algorithm~2 runtimes against theoretical $n^2$ (dashed blue) and $n \log n$ (dotted purple) scaling curves. All curves are normalized to the runtime for $\sim$50,000 galaxies. Algorithm~2 (black circles) scales significantly better than Algorithm~1 (red triangles), closely following $n \log n$ up to several hundred thousand galaxies. \textbf{Right panel:} Raw runtime of \texttt{Nessie} on the Setonix supercomputer, with an $n \log n$ fit (black line) shown up to two million galaxies. A departure from ideal scaling is visible beyond this point, likely due to system-level parallelization limits.
    }

    \label{fig:scalability}
\end{figure*}

We demonstrate the scalability of the improved group-finding algorithm implemented in \texttt{Nessie} (Algorithm~\ref{algo: improved algo}) in Fig. \ref{fig:scalability}. The left panel compares the runtime scaling of the original algorithm (Algorithm\ref{algo: 1}) and the new implementation, using a test dataset constructed by cloning the equatorial regions of the GAMA DR4 catalogue and redistributing them across the sky. This effectively increases the survey area while preserving local density and clustering structure. All theoretical scaling curves in this panel are normalized to the runtime of a single GAMA field.

It is clear from this comparison that while \texttt{Nessie}'s performance lies slightly above the ideal $\mathcal{O}(n \log n)$ curve, it remains significantly closer to this than to $\mathcal{O}(n^2)$, representing a substantial improvement over the original algorithm, which follows a clear $\mathcal{O}(n^2)$ scaling trend.

The right panel of Fig.~\ref{fig:scalability} shows the actual runtime of \texttt{Nessie} measured on a single node of the Setonix supercomputer at the Pawsey Supercomputing Centre. Up to two million galaxies, the runtime closely follows an $n \log (n)$ scaling, as indicated by the fitted black line. Beyond this threshold, however, we noticed a deviation from this relation, with the runtime increasing more steeply. We attribute this to a reduction in parallel efficiency as opposed to the algorithm itself, likely due to available cache sizes or thread management overheads on the specific system used.

Nonetheless, \texttt{Nessie} maintains near-optimal scaling across a broad range of input sizes and is capable of processing several million galaxies in just a few seconds. We emphasize that while such benchmarks offer valuable insight into algorithmic performance, the results are inherently dependent on the underlying hardware and software environment, and performance may vary between systems.

The FoF algorithm could be faster than this current implementation if one or both of the linking lengths were chosen to be constant. However, this would limit flexibility and arguably induce biases in the final group finding. It is also worth noting that we were able to push the algorithm to $\mathcal{O}(n \log n)$ complexity by implementing a simple spatial grid, where every cell had a width of twice the max on-sky linking length and then use a hash-map lookup to identify the neighbouring cells. However, because of redshift distortion, the redshift dependence on both $V_{\rm max}(z)$ and $D_{\rm max}(z)$, and the fact that the linking lengths between every pair of galaxies is unique to those pairs, we cannot mathematically guarantee that this solution will work for all galaxy surveys with any geometry. In order to keep \texttt{Nessie} as generalized as possible we have chosen not to implement this in version 1.0, however, later versions of \texttt{Nessie} could receive significant boosts exploring this area in more depth.

\subsubsection{Fearless Concurrency}
One of the core goals of \texttt{Rust} is to manage concurrent programming in a way that avoids many common errors and pitfalls, which are caught at compile time rather than at runtime. This enables users to easily implement multi-threading and parallelization without worrying about memory issues such as data races and deadlocks \citep{klabnik2018}.

We take full advantage of this capability by using the \texttt{rayon}\footnote{\url{https://crates.io/crates/rayon}} crate in \texttt{Rust} to parallelize the link-finding algorithm. This provides a straightforward speed-up, allowing \texttt{Nessie}'s performance to scale with the number of available cores. Because this parallelization is handled automatically, users benefit from improved performance without needing to restructure their code. Moreover, since the outer loop of the final link-finding algorithm (Alg. \ref{algo: improved algo}) is a simple \texttt{for} loop from $0..(n-1)$, where each iteration is independent, it can be perfectly parallelized. This means that \texttt{Nessie} can fully leverage high-performance computing environments, where many cores are available for processing.

\subsection{Building the group catalogue from links}

The original FoF algorithm described in \cite{huchra1982}, builds the galaxy groups directly whilst iterating over the redshift catalogue. In other words, it immediately begins to construct the group that galaxy $i$ belongs to, and so constructs a new group with every outer loop iteration. This is computationally inefficient and more modern implementations of the algorithm, and the one that we adopt for \texttt{Nessie}, first compute the pairwise links between all galaxies -- resulting in a list of tuples, as described in the previous section -- and then build an undirected graph where each node represents a galaxy and each edge corresponds to a valid link. Graph construction is typically very fast, thanks to modern graph libraries, which implement highly optimized data structures (e.g., adjacency lists, bitmaps) and bulk insertion methods. This method has been implemented in the GAMA group finder since 2018 and formed a core part of \texttt{pyFoF} \citep{lambert2020}.

Once the graph is built, the final group catalogue can be extracted by identifying connected components (i.e, subgraphs) in which every node is reachable from every other node. This step is highly optimized in graph libraries using traversal algorithms like Depth-First Search (DFS) and Breadth-First Search (BFS), which operate in linear time, $\mathcal{O}(V + E)$,
where $V$ is the number of vertices and $E$ is the number of edges \citep{morin2014}. This makes the overall process both efficient and scalable for large redshift surveys. We use the graph libraries \texttt{igraph}\footnote{\url{https://cran.r-project.org/web/packages/igraph/index.html}} and \texttt{NetworkX}\footnote{\url{https://github.com/networkx/networkx}} to perform this step in the \texttt{R} and \texttt{python} versions of \texttt{Nessie} respectively.

\section{Results}
\subsection{Tuning on GALFORM}
Since \texttt{Nessie} is built upon the GAMA group-finder in \cite{robotham2011}, then a key test is to retune the group finder on the same mock catalogue. An $S_{\rm total}$ score that is equivalent to \cite{robotham2011} using the same mock catalogue would imply that the changes made to the algorithm are appropriate, and that \texttt{Nessie} does at least as well as the original.

To do this we use the same mock catalogues as \cite{robotham2011}: a set of nine lightcones consisting of three GAMA fields each -- G09, G12, and G15 -- which were built from the Millennium simulation \citep{springel2005}, using the semi-analytical model GALFORM \citep{bower2006}. Their r-band magnitudes were abundance matched to the GAMA data to match the GAMA selection function and their magnitudes were adjusted with the same GAMA $k + e$-correction described in Equation 8 of \cite{robotham2011}. We use the mock catalogues that were made available in the first GAMA data release \citep{driver2011}, through the GAMA database schema browser.

We tuned the mock catalogue using the $R$ version of \texttt{Nessie} which has a built-in \texttt{tune\_group\_finder} function which wraps a function call to the $R$ \texttt{Highlander} package\footnote{\url{https://github.com/asgr/Highlander}}, a flexible genetic and Markov Chain Monte Carlo (MCMC) fitting framework. \texttt{Highlander} combines global and local optimization strategies by alternating between Covariance Matrix Adaptation (CMA) and the \texttt{LaplacesDemon} package (LD) for MCMC sampling. CMA is a global genetic algorithm based optimizer well-suited for exploring complex, multi-modal parameter spaces without requiring gradient information, while LD performs efficient posterior sampling using gradient-based MCMC algorithms such as Componentwise Hit-And-Run Metropolis (CHARM). We ran highlander using the two phases with the CMA/CHARM phase using 100 iterations and the MCMC phases using 2500.

We find best fit values $b_0=0.04\pm 0.01$ and $R_0 = 36 \pm 4$ with the best $S_{\rm total} = 0.38$. $b_0$ is close, but lower than the value of $b_0=0.06$ found in \cite{robotham2011} and the radial linking length is double the $R_0 = 18$. However, we don't expect these values to be identical to \cite{robotham2011} since 1) the algorithm has changed slightly with the replacement of the luminosity function with the mean density and should have some adjustments to the linking lengths and 2) \cite{robotham2011} used a simpler \texttt{Optim} function in \texttt{R}. More importantly, since the two linking lengths are coupled (the radial direction being a scaling of the plane-of-sky direction) a decrease in the plane-of-sky would naturally incur an increase in the radial direction. Furthermore, we notice that the algorithm can tolerate a large range of radial direction linking lengths without changing $S_{\rm total}$; $R_0$ can double in some cases and have a marginal effect on the score. More interesting is the value of $S_{\rm total}$ which is very close to the value of $S_{\rm total} = 0.41$ found in \cite{robotham2011}. This represents an average overlap of $\sim 78\%$. The fact that we can very closely recover the same tuning score as \cite{robotham2011}, albeit with different linking lengths, shows that the changes to the algorithm are reasonable.

\subsection{Comparison to GAMA}
To further test \texttt{Nessie} we compare \texttt{Nessie}'s results against the latest version of the GAMA group catalogue (G3Cv10). We download the galaxy catalogue that was used in the group finding in GAMA Data release 4\footnote{\url{https://www.gama-survey.org/dr4/schema/table.php?id=588}}. This includes the g02, G09, G12, and G15 regions. We excluded the g02 region due to low completeness in that field. This results in a final galaxy catalogue of 184 081 galaxies among the three GAMA equatorial regions. Using the ``GroupID'' column as ``truth'', we ran \texttt{Nessie} with the same $b_0$ and $R_0$ values as in \cite{robotham2011}, calculating the $S_{\rm total}$ score. We get a score of $S_{\rm total} \sim 0.6$ which is $\sim 90\%$ overlap. 

\begin{figure}
    \centering
    \includegraphics[width=\linewidth]{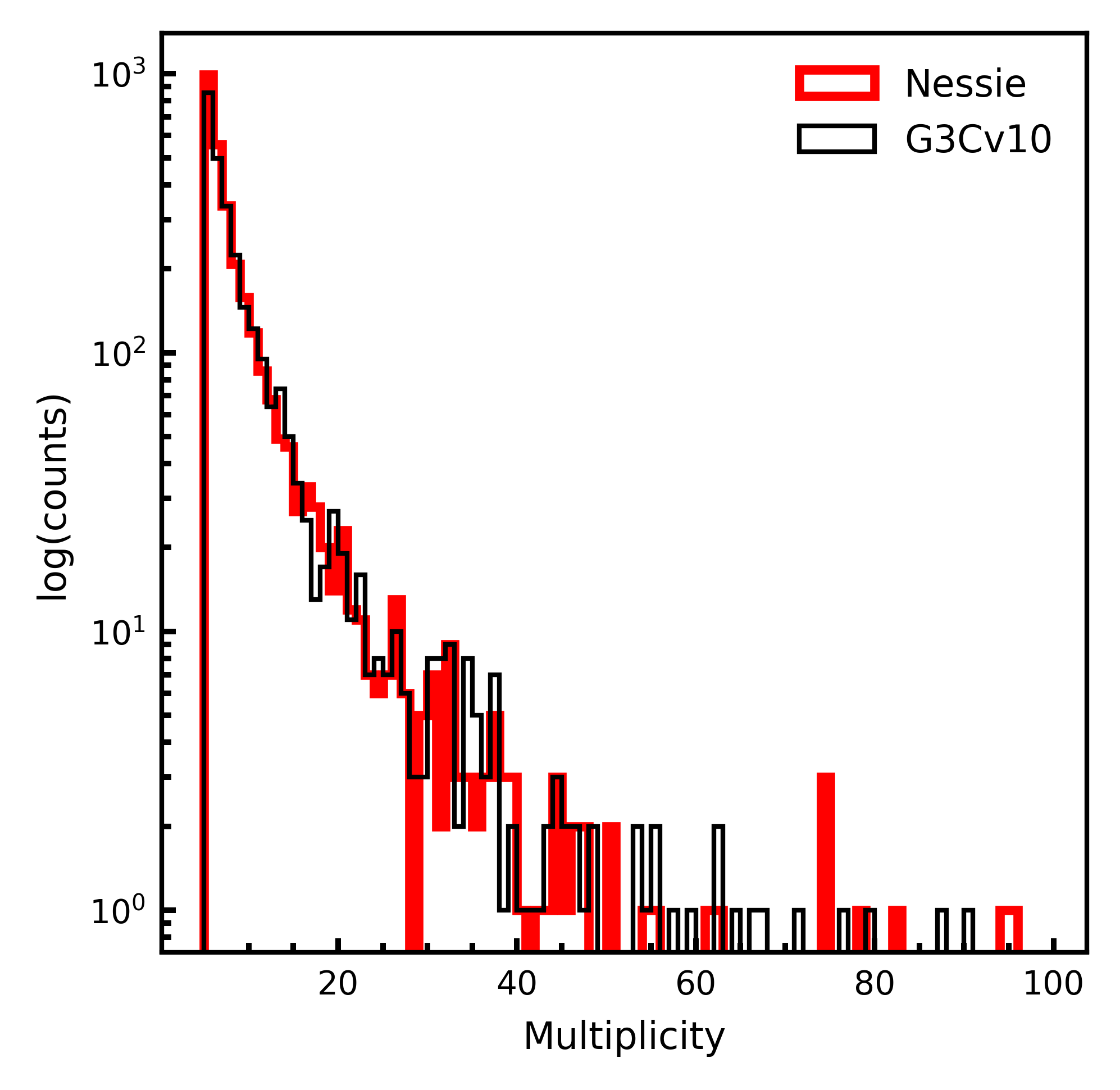}
    \includegraphics[width=\linewidth]{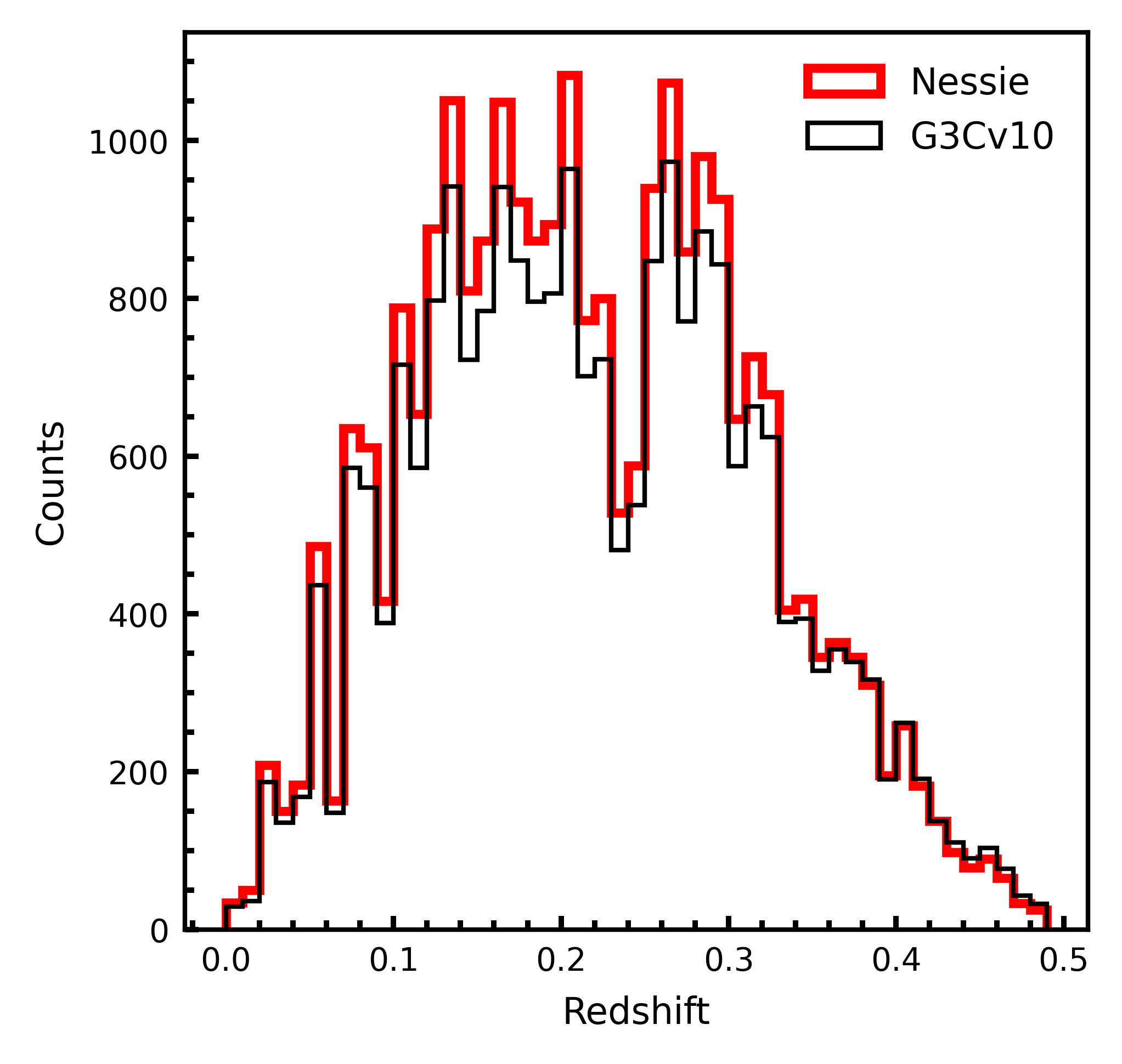}
        \caption{Distribution of multiplicity (top) and $n(z)$ (bottom) between G3Cv10 and \texttt{Nessie} using the same optimal parameters found in Robotham et al. (2011). }
    \label{fig:gama_group_nz}
\end{figure}

In addition, we also compare the multiplicity and the $n(z)$ distributions in Figure \ref{fig:gama_group_nz}. The median redshift and multiplicity were recalculated using the same method to ensure that any differences are purely due to the group-finding algorithm. Both of the distributions lie nearly on top of one another. Although it should be noted that \texttt{Nessie} did find slightly more groups in the middle redshift range of the survey ($0.05 \leq z \leq 0.35$); these groups are small two and three member groups. Overall agreement between both catalogues bodes incredibly well for \texttt{Nessie} as it suggests that the algorithm is behaving in much the same way as the previous renditions of the GAMA group-finder.

\begin{figure}
    \centering
    \includegraphics[width=0.48\linewidth]{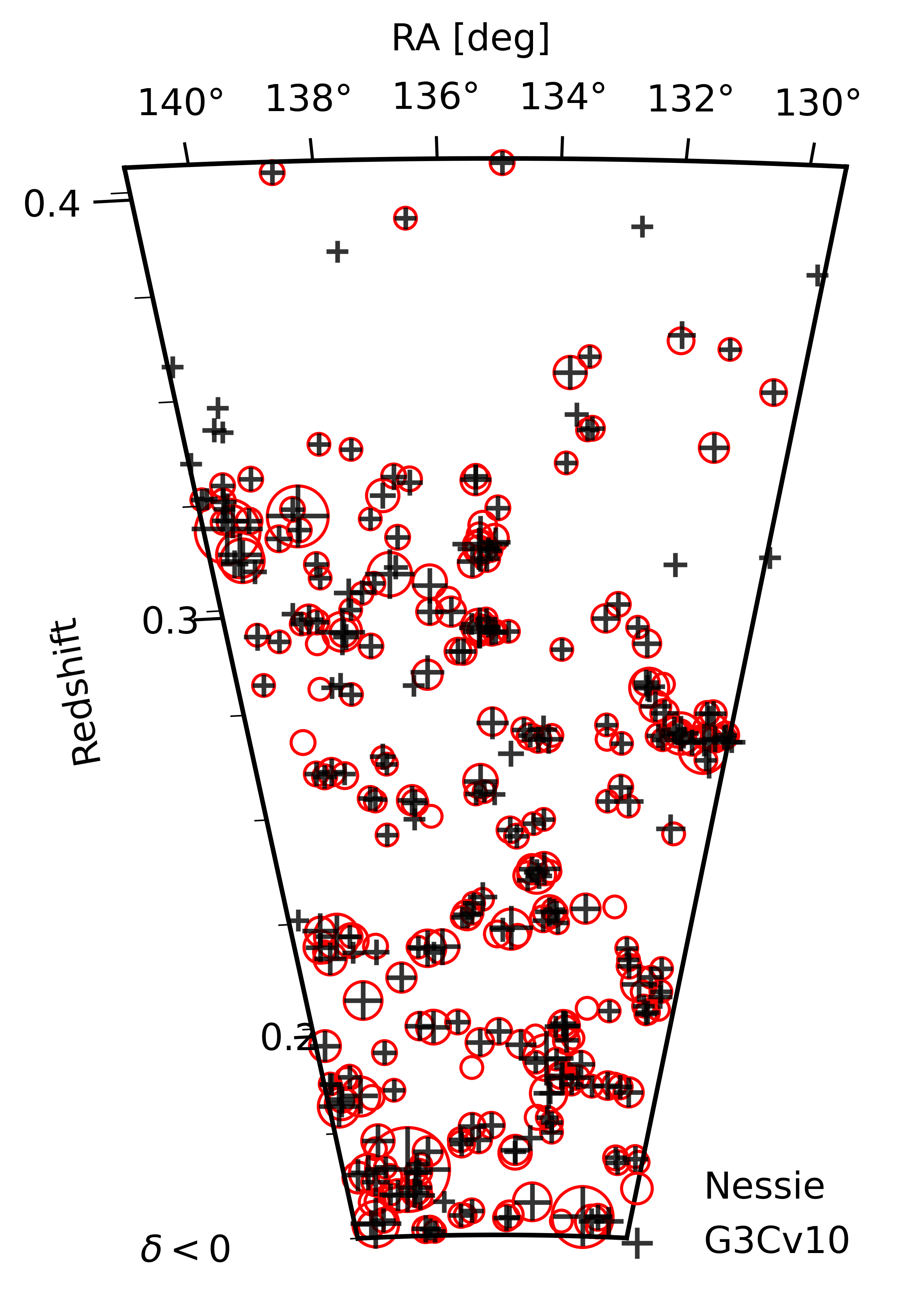}
    \includegraphics[width=0.48\linewidth]{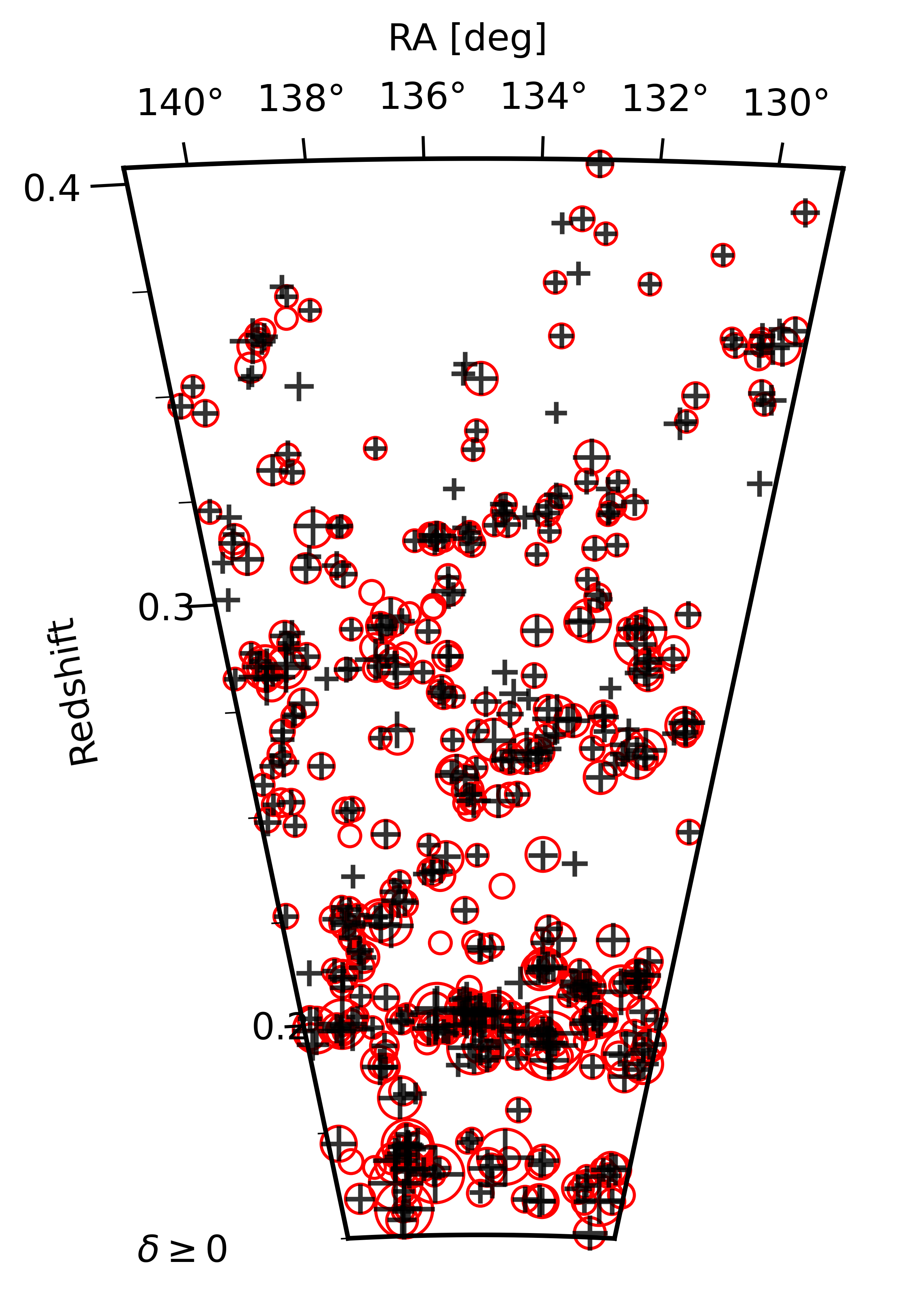}
    \caption{Three-dimensional distribution of galaxy-groups with five or more members in the G3Cv10 group catalogue (black crosses) and \texttt{Nessie} (red circles), in the redshift range $0.15 < z < 0.4$ for the G09 region.}
    \label{fig:middle_g09_region}
\end{figure}

\begin{figure}
    \centering
    \includegraphics[width=0.48\linewidth]{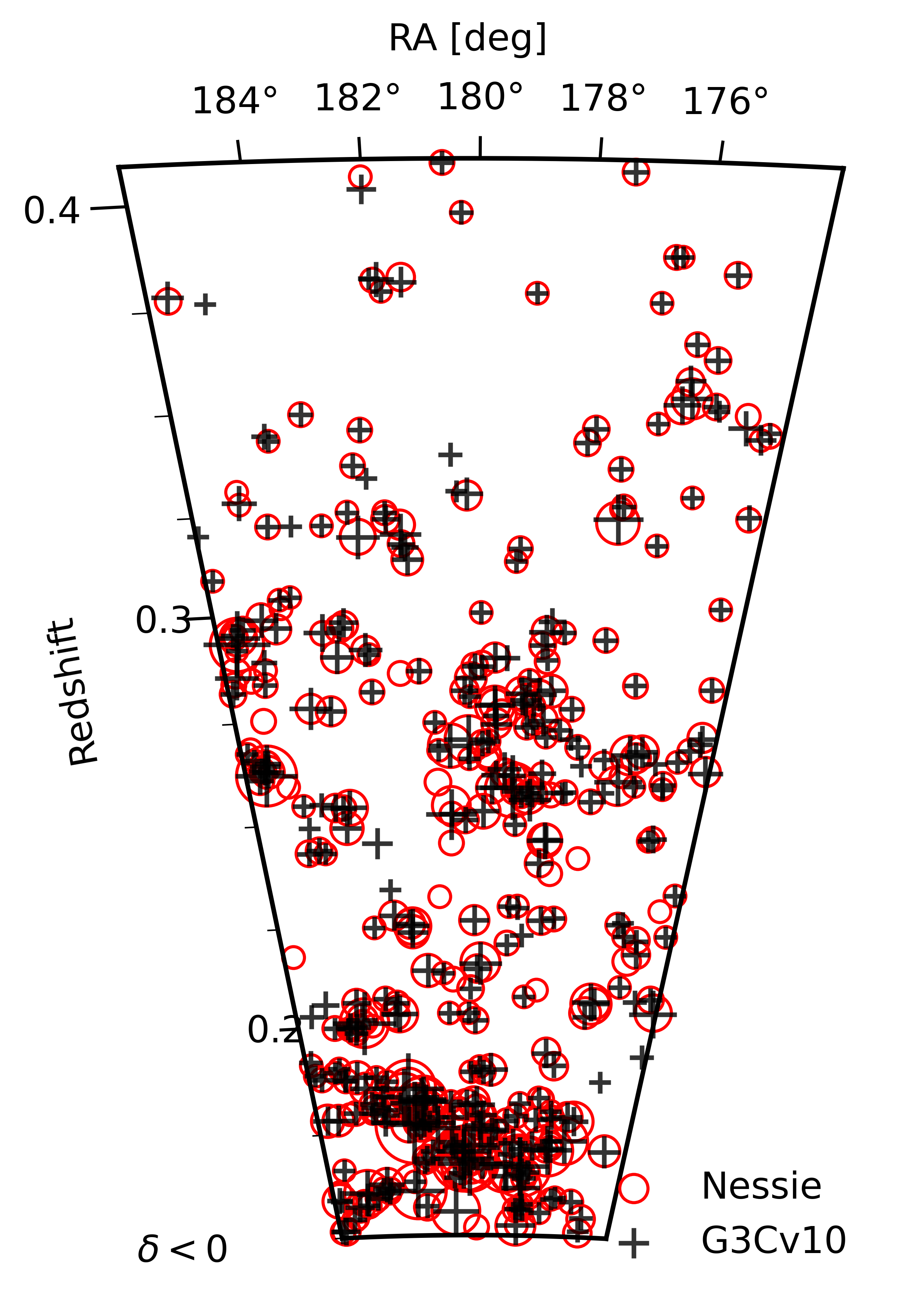}
    \includegraphics[width=0.48\linewidth]{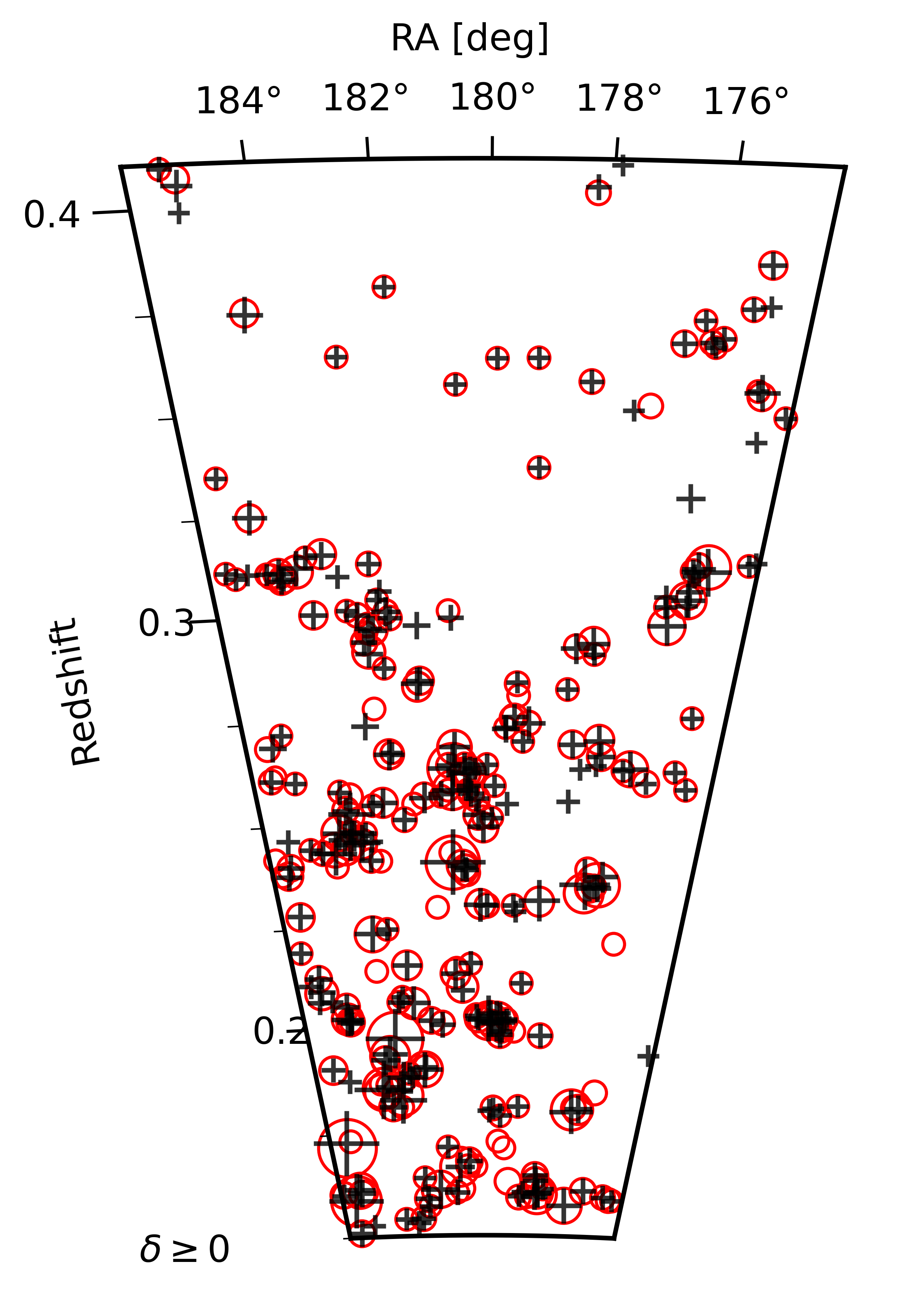}
    \caption{Same as Fig. \ref{fig:middle_g09_region} but showing the G12 region.}
    \label{fig:middle_g12_region}
\end{figure}

\begin{figure}
    \centering
    \includegraphics[width=0.48\linewidth]{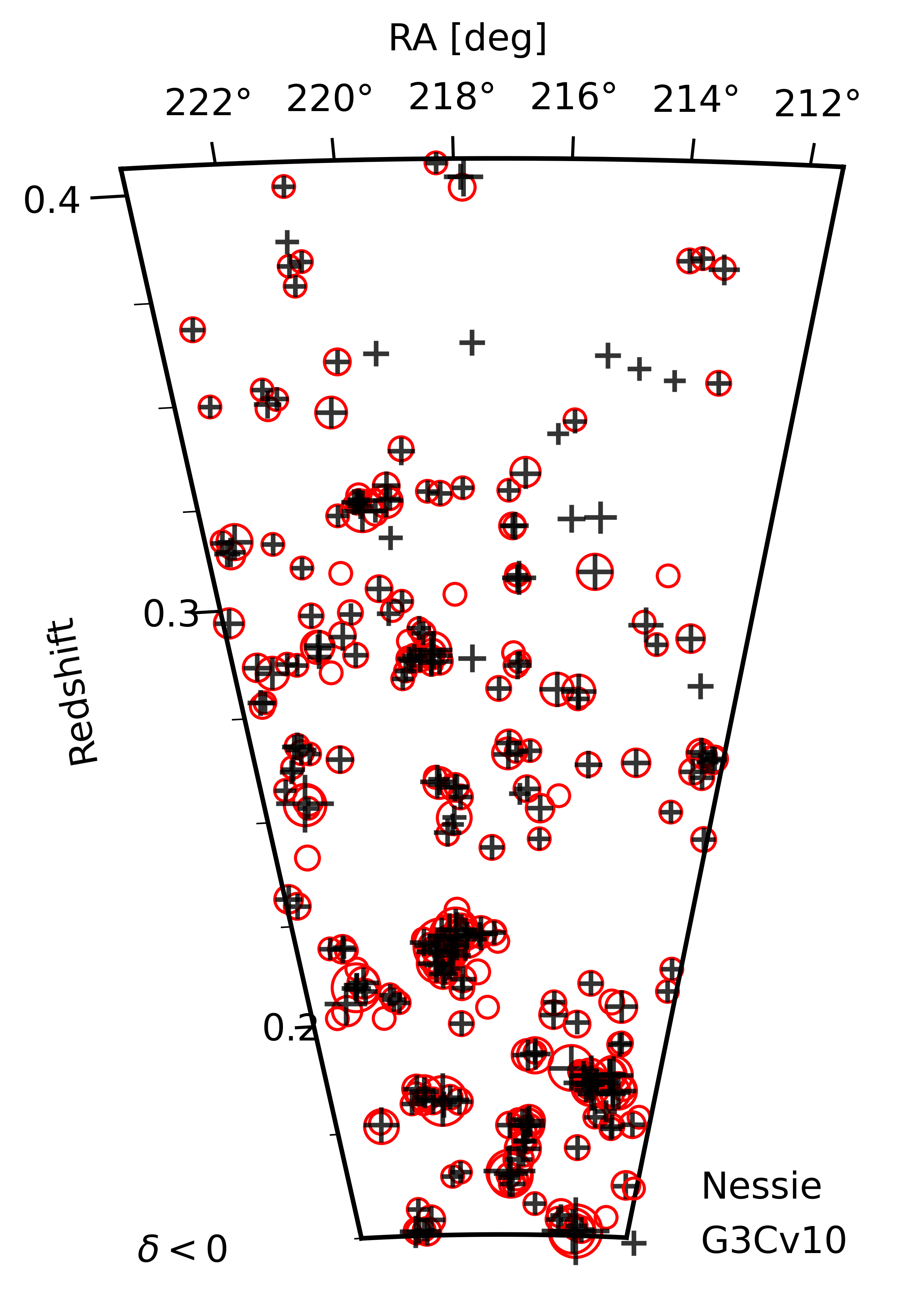}
    \includegraphics[width=0.48\linewidth]{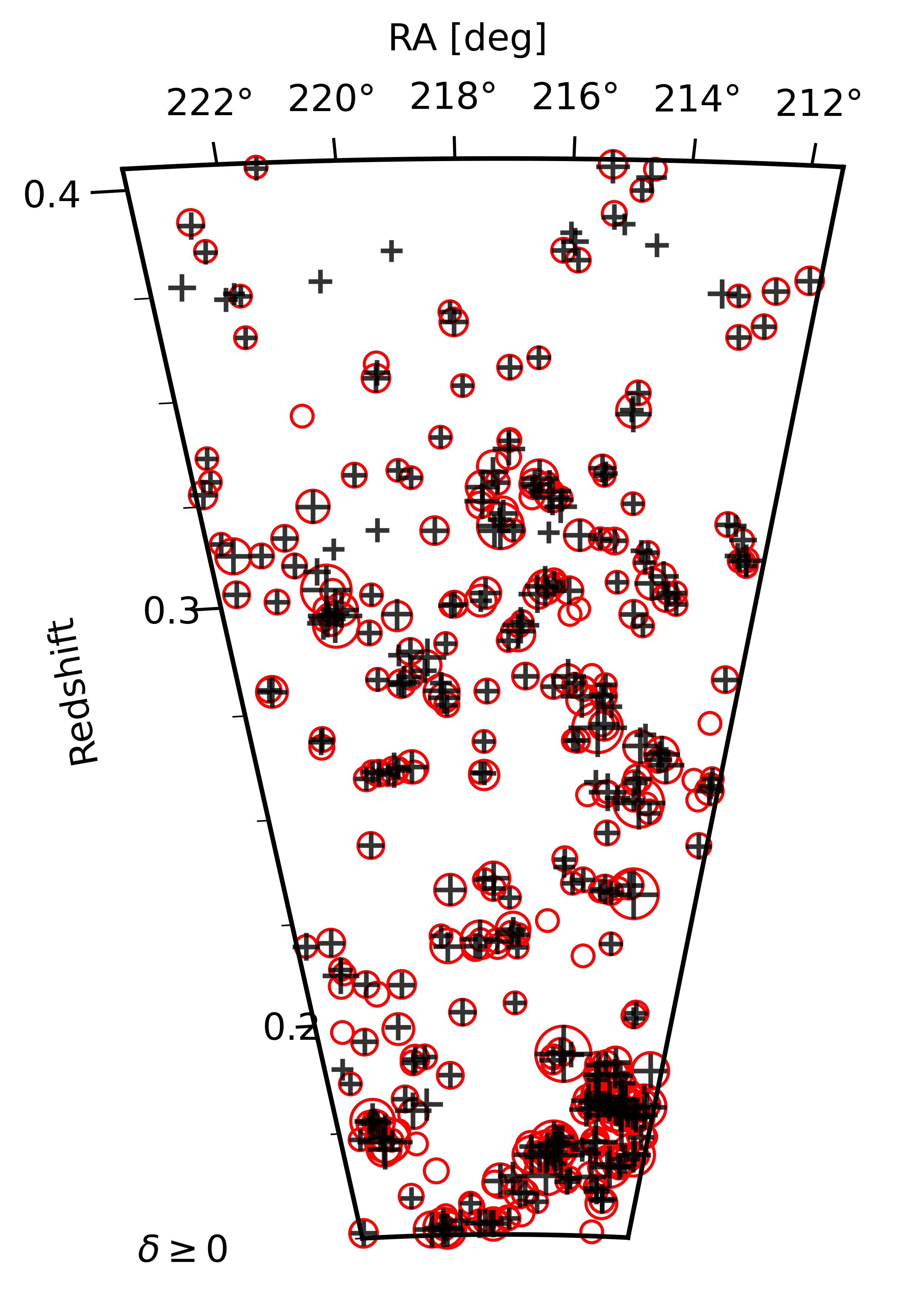}
    \caption{Same as Fig. \ref{fig:middle_g09_region} but showing the G15 region.}
    \label{fig:middle_g15_region}
\end{figure}

We also inspect the positional information of both catalogues by examining slices of both catalogues across the GAMA regions G09, G12, and G15 in Figs. \ref{fig:middle_g09_region}, \ref{fig:middle_g12_region}, and \ref{fig:middle_g15_region}. All these Figures are between 0.15 and 0.4 in redshift, and split between negative and positive declination to make the overlap between the two catalogues easier to see. The red circles are the groups found by \texttt{Nessie}. The black disks are the G3Cv10 catalogue. The points have been scaled according to the number of members. 

Remarkably, there is nearly a perfect overlap between the two catalogues. There are some smaller groups that are found in \texttt{Nessie} which are not found in G3Cv10. Of the overlapping groups both their position, and number of members are, for the most part identical. All the large-scale structure, in particular, is recovered exactly. 

\subsection{Comparison to SDSS}
To test the generalized nature of \texttt{Nessie}, we run it on the Sloan Digital Sky Survey (SDSS) Data Release 12 \citep{eisenstein2011,alam2015,kollmeier2019} using a very bare-bones set up. We compare \texttt{Nessie} to the galaxy group catalogue that was built on the Legacy Survey in \citep{tempel2017}. This survey spans a large range in declination from $-3.8 < \delta < 70.2$, up to a redshift of $0.2$. This is very different to the equatorial regions of GAMA in survey design and consists of 584 449 galaxies (nearly three times more than GAMA DR4).

The galaxy catalogue that \cite{tempel2017} ran their group-finder on is publicly available from the \textit{Vizier} online portal. From \textit{Vizier}, we download the RA, Dec, redshift, and the group ID of all the galaxies.

We performed an out-of-the-box test using the bare minimum information. We generate $\bar{\rho}(z)$ using the helper function \texttt{create\_density\_function} in \texttt{Nessie} which applies the method described in Eqs. \ref{eq: integral} and \ref{eq: bins} and choosing $n(z)$ to simply be the $n(z)$ measured from the data itself. While this is a good first order approximation to the $n(z)$, a more robust measure (using the methods described in \cite{cole2011} or \cite{farrow2015} for example) is required for to be more scientifically rigorous. 
After building the density function we take the \cite{tempel2017} group catalogue to be ``truth'' and perform a very simplistic tuning using the \texttt{optimize$\_$nm} function in \texttt{Nessie} which performs a Nelder-Mead algorithm to find the optimum values using the same cost function described in Section 2.2, namely $S_{\rm tot}$.

Even with this elementary approach, we were able to achieve an $S_{\rm total} = 0.5$ which implies about an 84$\%$ overlap with \cite{tempel2017}.

\begin{figure}
    \centering
    \includegraphics[width=\linewidth]{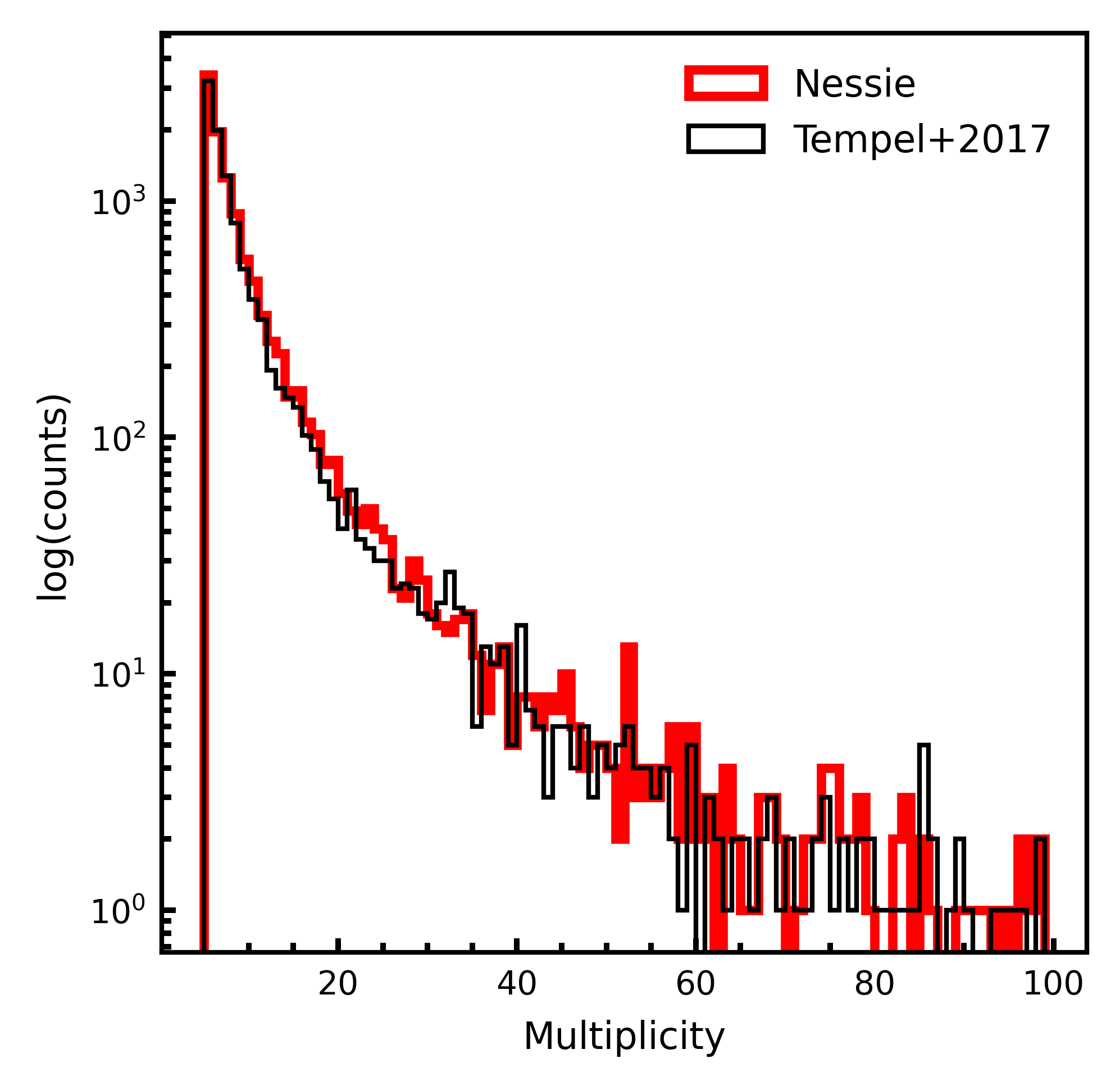}
    \includegraphics[width=\linewidth]{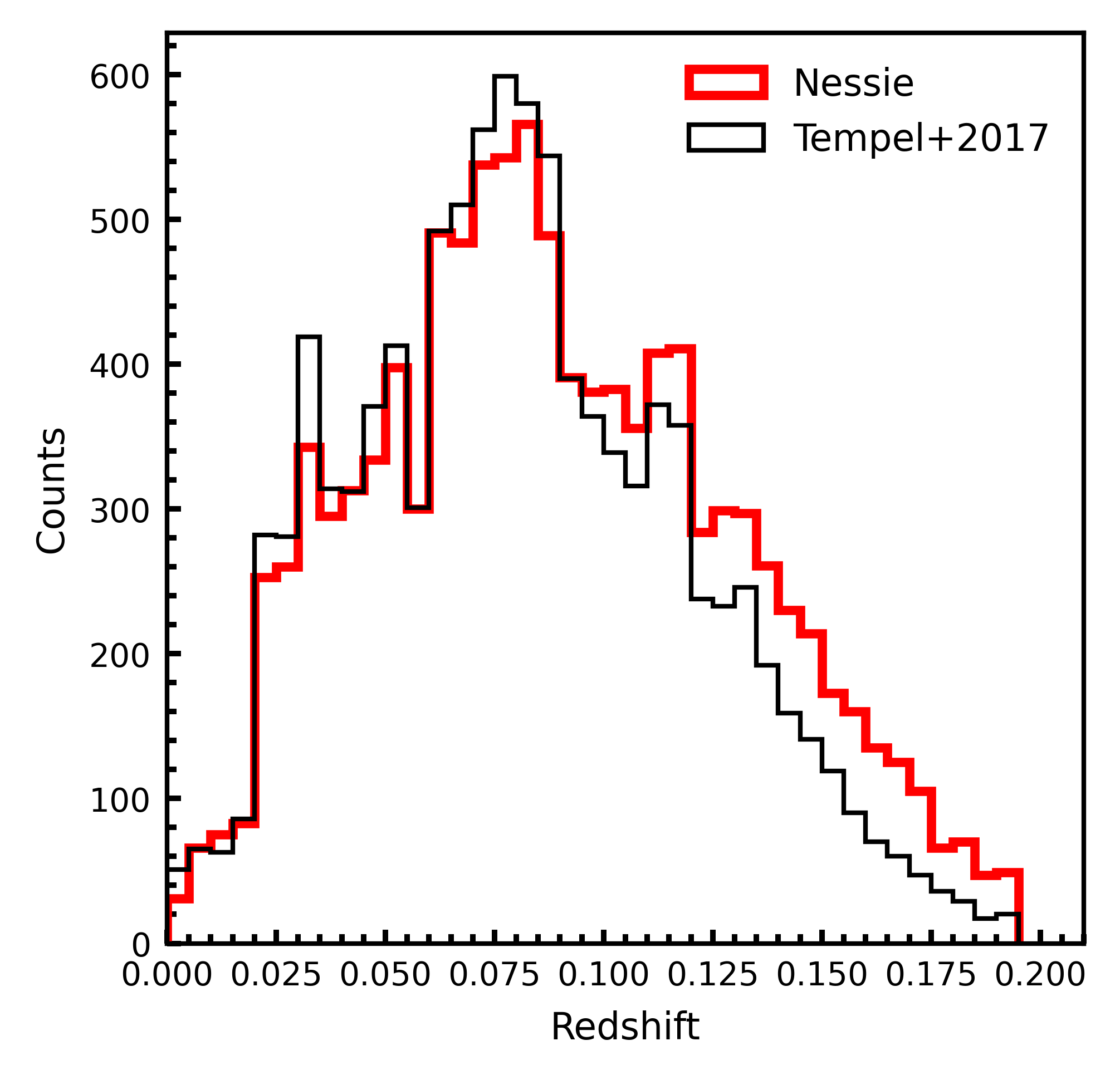}
    \caption{Comparison between the multiplicity distribution (top) and the $n(z)$ (bottom) of galaxy-groups in the SDSS group catalogue and \texttt{Nessie}}
    \label{fig:sdss_n_of_z}
\end{figure}

In addition, we compared the distributions of the multiplicity and the $n(z)$ distribution of the group catalogue published in \cite{tempel2017} to the catalogue generated from \texttt{Nessie}. Again, we standardized the method for calculating the group-properties across both catalogues. The multiplicity distribution in Fig. \ref{fig:sdss_n_of_z} is very closely aligned and the $n(z)$ distribution has the same overall shape, although \texttt{Nessie} did find fewer groups within $0.03 < z < 0.1$ and slightly more at $z > 0.1$.

However, we emphasize that the goal of this comparison is not to perfectly build a SDSS catalogue, but simply show that \texttt{Nessie} gets similar results with very minimal effort. Naturally, when building a scientific catalogue much more effort would need to go into building the density function and finding the optimum set of parameters. In our testing we note that getting the correct $\bar{\rho}(z)$ is critical. However, it is quite remarkable that such a strong agreement could easily be found between two entirely independent methods and speaks to the generalized nature of \texttt{Nessie}, showing that the algorithmic changes are reasonable.

\section{Discussion}

\subsection{Verifying \texttt{Nessie} Group Catalogues}

Tuning on the GALFORM mock catalogues suggests that \texttt{Nessie} performs at least as well as previous versions of the GAMA group catalogue. More importantly, the strong agreement between our \texttt{Nessie}-based catalogue and the 10th version of the GAMA group catalogue—constructed using a fiducial parameter selection—justifies the simplifications we introduced.

Setting $E_b = E_R = 0$ was already well motivated by the findings of \citet{robotham2011}, but our results further demonstrate that the dependence of linking lengths on intrinsic galaxy brightness is negligible. Setting $\nu = 0$ not only simplifies the algorithm conceptually but also reduces computational complexity, especially during the tuning phase on mock catalogues. Furthermore, it is worth noting that this choice seems appropriate for different survey geometries including pencil-beam like surveys such as the Deep Extragalactic VIsible Legacy Survey \citep[DEVILS;][]{davies2018}; when tuning the full GAMA group-finding algorithm on DEVILS-like lightcones, built using Shark v2 simulations \citep{lagos2024}, Bravo et al. (\text{in prep}) found optimum $E_b$ and $E_r$ values of -0.08, and 0.12 respectively which are very near to 0 and have a negligible effect on the individual linking length scaling. It is also worth noting that they also set $\nu = 0$. This sample is vastly different to both the GAMA and SDSS samples used here, covering a six square-degree field, out to a redshift of 1. This once again justifies our choice of setting $E_b = E_r = 0$.

The negligible difference in the $S_{\rm total}$ score between the original GAMA implementation and \texttt{Nessie} indicates that our algorithm retains the core functionality of the established method. Together, the mock tuning and the comparison to the latest GAMA catalogue provide substantial evidence that \texttt{Nessie} functions as intended.

Beyond GAMA, our application to SDSS data further highlights \texttt{Nessie}'s adaptability. Despite being developed and tested primarily using GAMA data, \texttt{Nessie} successfully generated a group catalogue for SDSS—an entirely different survey in terms of depth, area, and sky footprint—using only positional and redshift information. This catalogue had a nearly $85\%$ overlap with \cite{tempel2017} and was built in a fraction of the time. This serves as a strong demonstration of the algorithm’s generalizability.

Frequently, running group-finders on new surveys often entails significant overhead, such as adjusting for different magnitude limits and constructing survey-specific luminosity functions to determine linking lengths. However, our analysis with SDSS shows that \texttt{Nessie} can approximate other group finders with minimal input, reinforcing its flexibility and practicality.

The aim of this paper is not to construct a definitive group catalogue, but rather to present a robust and generalizable tool for doing so. The final catalogue quality will ultimately depend on user choices—particularly in defining the $n(z)$ distribution and selecting optimal values for $b_0$ and $R_0$. Some users may choose to calibrate these parameters using simulations (depending on which they trust), while others may rely on prior calibrations from existing surveys. In either case, a carefully chosen set of parameters should yield high-quality catalogues for any redshift survey, regardless of its geometry.

Moreover, constructing the SDSS group catalogue using only the tools provided by \texttt{Nessie} proved straightforward, with minimal code overhead. Most user effort lies in pre-processing and plotting data. This ease of use is a major strength—reducing user-side coding lowers the chance of bugs or implementation errors and streamlines the overall workflow.

\subsection{Group Shredding and Over-clustering}

Over-clustering (where group finders erroneously connect groups together that should be separate) and shredding (where large groups and clusters are instead found to be many smaller groups) are difficult problems to solve and usually require post-processing \citep[e.g.][]{tempel2016} which will be the focus of future versions of \texttt{Nessie}. However, we comment briefly on these effects in using the GALFORM mock catalogues.

\texttt{Nessie} does well to recover the core of the larger mock groups, often recovering $\sim 90\%$ of group members correctly with only $~\sim 3\%$ of the group members closest to the edge of the group being incorrectly identified as isolated. The remaining members are found to be in smaller groups that happen to also be at the edges of the core of the main group. From this it seems that \texttt{Nessie} does well in recovering real large groups and avoiding shredding.

We also looked at the inverse effect by examining large groups that were found by \texttt{Nessie}. About $\sim 75\%$ of the group members were indeed correctly identified within the group with $\sim 15\%$ of the group members actually being isolated galaxies. These false group members all tend to be close to the main part of the group which would make correctly identifying them as isolated incredibly difficult for any FoF algorithm. The remaining false group members are actual groups that have been incorrectly assigned to the larger group. From these examples it seems that \texttt{Nessie} has a slight tendency to over-cluster as opposed to shredding and future work to identify erroneous links might be needed in the future.

Both shredding and over-clustering can be managed by changing the values of the linking lengths. Indeed, the payoff between over-clustering and shredding is part of the tuning process and is accounted for within the $S_{\rm tot}$ score. It's possible that a different statistic would result in different over-clustering and tuning behaviour. In the extreme cases, we could choose purity as the optimizing static which would result in the linking lengths being very strict, but ruining completeness, and the opposite effect would be seen if we chose completeness as the optimizing statistic instead. Choosing to optimize using $S_{\rm tot}$ seems to slightly favour completeness over purity resulting in less shredding but more over-clustering.

These conclusions are based off of a very basic, qualitative analysis on one mock catalogue for a particular survey using a single metric. A much more complete and nuanced analysis would be required to fully characterize these effects and will form part of later work to incorporate post-processing steps to improve membership assignment.

\subsection{Speed and Scalability}

Significant effort has gone into making \texttt{Nessie} as fast and scalable as possible. Many of the algorithmic improvements—such as converting angular to Euclidean searches, employing binary search for line-of-sight selection, and utilizing a k-d tree for plane-of-sky selection—can benefit any implementation of the FoF algorithm, regardless of language or survey. These optimizations substantially improve the base algorithm's performance.

However, our primary goal was to deliver a tool that is not only algorithmically efficient but also ready for the next generation of redshift surveys. \texttt{Nessie} achieves this through a combination of fast data structures, implementation in a high-performance language (\texttt{Rust}), and full multi-threading.

On a MacBook Pro with 32 GB RAM and an Apple M3 Pro chip (11 cores), \texttt{Nessie} processed the three GAMA equatorial regions (180k galaxies) in approximately 0.1 seconds. The SDSS catalogue, consisting of 584k galaxies, was processed in under 4 seconds and for tests on mock surveys containing $\sim 1$ million galaxies, \texttt{Nessie} completed group finding in less than 10 seconds.

Of course, these benchmarks are system-dependent and sensitive to algorithm settings—e.g., increasing linking lengths increases runtime. Nonetheless, they demonstrate what is possible on a reasonably standard setup. Because \texttt{Nessie} scales linearly with core count, doubling available cores should roughly halve the runtime, offering clear benefits on more powerful systems. 

These benchmarks highlight the fact that \texttt{Nessie} is already capable of handling the data volumes we expect in the next generation of redshift surveys.

\subsection{Accessibility}

We have designed \texttt{Nessie} to be as accessible and user-friendly as possible, offering interfaces in both \texttt{Python} and \texttt{R}. Installation is straightforward via \texttt{pip} for \texttt{Python} users and \texttt{devtools} for \texttt{R} users. All packages—including the core \texttt{Rust} back-end, \texttt{fof}—maintain over 80\% test coverage and are accompanied by thorough documentation and usage examples hosted on their respective GitHub repositories.

Running \texttt{Nessie} requires only the definition of $\bar{\rho}(z)$ (for which helper functions are provided), along with loading RA, Dec, and redshift values, and specifying the linking lengths. This clean, high-level interface reduces the potential for user error and streamlines the group-finding process, allowing users to focus on their science rather than implementation details.

We actively maintain \texttt{Nessie} on GitHub, where users are encouraged to submit bug reports, raise issues, and contribute to ongoing development. Contribution and bug reporting guidelines are outlined clearly in each repository.

To support new users, we provide several example workflows—including Jupyter notebooks—that walk through the process of building a group catalogue for surveys such as GAMA.

\section{Conclusions}

In this paper, we presented \texttt{Nessie}, a FoF group finder based on the algorithm used in the GAMA survey. Our primary goals were to develop a group finder that is easy to use, computationally efficient, and generalizable across a range of redshift surveys. To achieve these objectives, we:

\begin{itemize}
\item[1)] Developed an easily installable and user-friendly package in both \texttt{R} and \texttt{Python};\
\item[2)] Implemented algorithmic simplifications to reduce the number of expensive friend-checks, and multithreaded the core algorithm using \texttt{Rust};\
\item[3)] Eliminated the dependence on luminosity functions by defining linking lengths based on the redshift distribution, $n(z)$, which can be constructed directly from the dataset.
\end{itemize}

We verified that \texttt{Nessie} behaves as expected and closely reproduces the GAMA group catalogues. When applied to the same mock catalogues used in \citet{robotham2011}, it achieves a comparable $S_{\rm total}$ score. To test its generalizability, we also applied \texttt{Nessie} to a completely independent dataset—SDSS—and found an $85\%$ group overlap without any tuning.

The \texttt{Python} package\footnote{\url{https://github.com/TrystanScottLambert/nessie_py}} is available via \texttt{pip}, and the \texttt{R} package\footnote{\url{https://github.com/TrystanScottLambert/Nessie}} can be installed directly from GitHub using \texttt{devtools}. Both implementations are fully documented and thoroughly tested.

\subsection{Future work}
Future versions of \texttt{Nessie} will likely become even more optimized, as there is still considerable work to be done—particularly in exploring optimal data structures (e.g., octrees, HEALPix tessellation) or even developing new ones. We have already mentioned that we were able to get $\mathcal{O}(n\log n)$ performance by using a simple spatial grid, and if this can be shown to work universally this could greatly improve speed. Furthermore, techniques exploiting hardware such as GPU acceleration might be reasonable avenues of exploration. Any such improvements will propagate to both the \texttt{Python} and \texttt{R} packages simply by bumping the version of the \texttt{fof} dependency.

Several post-processing steps—such as identifying over- or under-clustering and applying appropriate corrections (e.g., the method in \cite{tempel2016})—remain to be implemented. Incorporating such post-processing functionality would not only enhance the creation of new group catalogues but also enable the refinement of legacy ones.

The most complicated aspect of running \texttt{Nessie} is building the $\bar{\rho}(z)$ function and our testing shows that this is a critical step. Although \texttt{Nessie} does provide a helper function to construct $\bar{\rho}(z)$, in the future it would be much more user-friendly to incorporate this construction directly from the redshifts. We plan on using the routine in \cite{cole2011} for constructing the $n(z)$ function in a way that removes the underlying large-scale structure, and using this to construct the density function. This would make \texttt{Nessie} far more user-friendly. However, we provide the user with the helper function in this initial release which allows for the maximum flexibility in determining the density function.

\section*{Acknowledgements}
The authors would like to thank the two anonymous reviewers for their feedback and suggestions.
TSL and SD acknowledge support from the Australian Research Council (ARC) Laureate Fellowship scheme (FL220100191). ASGR acknowledges funding by the ARC Future Fellowship scheme (FT200100375). CL acknowledges funding via the ARC Discovery Project DP210101945. This work was supported by resources provided by the Pawsey Supercomputing Centre with funding from the Australian Government and the Government of Western Australia.

%%%%%%%%%%%%%%%%%%%%%%%%%%%%%%%%%%%%%%%%%%%%%%%%%%
\section*{Data Availability}

%The inclusion of a Data Availability Statement is a requirement for articles published in RASTI. Data Availability Statements provide a standardized format for readers to understand the availability of data underlying the research results described in the article. The statement may refer to original data generated in the course of the study or to third-party data analysed in the article. The statement should describe and provide means of access, where possible, by linking to the data or providing the required accession numbers for the relevant databases or DOIs.

Nessie is publicly available under the MIT licence for both \texttt{R} (\url{https://github.com/TrystanScottLambert/Nessie}) and \texttt{Python} (\url{https://github.com/TrystanScottLambert/nessie_py}). All code used to generate the plots in this paper, as well as any non readily-available public-data  are made publicly available at \url{https://github.com/TrystanScottLambert/nessie_plots}. The GAMA and SDSS data used throughout this paper are publically available through \textit{Vizier} and the GAMA website respectively. The latex project to generate this paper is available upon reasonble request to the corresponding author.

\appendix

\section{Properties Overview}

\texttt{Nessie} outputs a range of core properties for galaxy groups, galaxy pairs. These properties are derived from \cite{robotham2011} and are designed to be close to publication-ready. While many properties are straightforward, a few—particularly those related to mass and luminosity—require further processing before use in scientific analysis or publication.

Since the provided properties are subject to change in the future it is worth reading the most recent \texttt{Nessie} documentation to get the full list of available group properties including units and definitions.

\subsection{Group Properties}

Each identified group is assigned a unique group ID and a multiplicity (i.e., the number of member galaxies). A range of structural, positional, and physical properties is also provided.

\subsubsection{Group Centre Definitions}

There is no unique definition for a group centre, so we include three common options:

\begin{itemize}
    \item[-] Brightest Central Galaxy (BCG)
    \item[-] Iterative centre
    \item[-] Flux-weighted centre
\end{itemize}

Each definition yields RA, Dec, and redshift coordinates. For the BCG and iterative centres, the index of the corresponding galaxy is also provided.

Following \cite{robotham2011}, we recommend the iterative centre as the most reliable estimator of the true group centre.

\subsubsection{Radial Measures}

As with group centres, there is no universally preferred radial measure. We adopt several projected radius definitions from \cite{robotham2011}, including:

\begin{itemize}
    \item[-] $r_{50}$ — radius enclosing 50\% of group members
    \item[-] $r_{100}$ — radius enclosing 100\% of group members
    \item[-] $r_{\sigma}$ — radius enclosing 66\% of group members
\end{itemize}

These radii are projected on the sky and help quantify the compactness of the group.

\subsubsection{Mass Estimation}

Group masses are estimated via the velocity dispersion using the "gapper" method \citep{beers1990} and applying the virial theorem. Both the velocity dispersion and virial mass are included as output properties.

However, users should treat these masses as lower limits. Scaling is typically required, especially to account for incompleteness due to magnitude limits. The correction factor $F$ accounts for undetected group members via the luminosity function $\Phi(M)$:

\begin{equation}
    F  = \frac{\int\limits_{-\infty}^{\infty}\Phi\left(M\right)dM}{\int\limits_{-\infty}^{M_{\rm lim}}\Phi\left(M\right)dM} ,
\end{equation}

where the absolute magnitude limit $M_{\rm lim}$ is computed as:

\begin{equation}
    M_{\rm lim} = m_{\rm lim} - 5\log(D_{c}) + 25,
\end{equation}

where $m_{\rm lim}$ is the survey's apparent magnitude limit and $D_c$ the group’s comoving distance.

The scaled total group mass is then:

\begin{equation}
    M_{\rm tot} = M_{N} F,
\end{equation}

where $M_N$ is the unscaled mass returned by \texttt{Nessie}. For more details, see Section 4.3 of \cite{robotham2011}.

\subsubsection{Brightness and Luminosity}

The total group luminosity proxy ($f_p$) is calculated by summing the fluxes of the group members. This value can also be scaled by the same factor $F$ used for mass correction.

The group’s total absolute magnitude is then estimated as:

\begin{equation}
    M = -2.5\log\left(f_p\right)
\end{equation}

Further discussion on flux scaling can be found in Section 4.4 of \cite{robotham2011}.

\subsection{Pair Properties}

We separately flag groups with a multiplicity of two as galaxy pairs. These are included in the main group catalogue but are also provided with a minimal set of dedicated properties for convenience.

\subsubsection{Position}

The iterative centre is less appropriate for pairs, as it defaults to one of the two galaxies. Instead, we use the flux-weighted position (RA, Dec, and redshift) as the preferred estimator.

\subsubsection{Total Magnitude}

Computed identically to the group case, using the combined flux of the two galaxies.

\subsubsection{Separations}

We provide both the projected angular separation between the two galaxies and the absolute difference in their redshifts.

\subsubsection{Identifiers}

Each pair retains the same group ID as in the group catalogue. The indices of both member galaxies are also included.

\bibliographystyle{elsarticle-harv} 
\bibliography{references}



%% For example:

%% \bibitem[Aladro et al.(2015)]{Aladro15} Aladro, R., Martín, S., Riquelme, D., et al. 2015, \aas, 579, A101

%%\end{thebibliography}

\end{document}